\documentclass[12pt]{article}
\newcommand{\be}{\begin{equation}}
\newcommand{\ee}{\end{equation}}

\def\bsg{\ifmmode B\to X_s\gamma\else $B\to X_s\gamma$\fi}
\def\bsll{\ifmmode B\to X_s\ell^+\ell^-\else $B\to X_s\ell^+\ell^-$\fi}
\def\shat{\ifmmode \hat{s}\else $\hat{s}$\fi}

\newcommand{\newc}{\newcommand}

\newc{\gsim}{\lower.7ex\hbox{$\;\stackrel{\textstyle>}{\sim}\;$}}
\newc{\lsim}{\lower.7ex\hbox{$\;\stackrel{\textstyle<}{\sim}\;$}}
\newc{\ie}{{\it i.e.}}
\newc{\etal}{{\it et al.}}
\newc{\mev}{\hbox{\rm\,MeV}}
\newc{\gev}{\hbox{\rm\,GeV}}
\newc{\tev}{\hbox{\rm\,TeV}}
\newc{\xpb}{\hbox{\rm\, pb}}
\newc{\xfb}{\hbox{\rm\, fb}}

\newc{\mtop}{m_t}
\newc{\mbot}{m_b}
\newc{\mz}{M_Z}
\newc{\mw}{M_W}
\newc{\alphasmz}{\alpha_s(M_Z)}
\newc{\swsq}{\sin^2\theta_W}
\newc{\cwsq}{\cos^2\theta_W}
\newc{\tw}{\tan\theta_W}
\newc{\cw}{\cos\theta_W}
\newc{\sw}{\sin\theta_W}
\newc{\BR}{\hbox{\rm BR}}
\newc{\zbb}{Z\to b\bar}
\newc{\Gb}{\Gamma (Z\to b\bar b)}
\newc{\Gh}{\Gamma (Z\to \hbox{\rm hadrons})}
\newc{\sgn}{\mbox{sgn}}

\newlength{\myem}
\newcounter{mysubequation}[equation]

\newcommand{\GeV}{\,\mathrm{GeV}}

\def\beq{\begin{equation}}
\def\eeq{\end{equation}}
\def\bea{\begin{eqnarray}}
\def\eea{\end{eqnarray}}
\def\slashchar#1{\setbox0=\hbox{$#1$}           
   \dimen0=\wd0                                 
   \setbox1=\hbox{/} \dimen1=\wd1               
   \ifdim\dimen0>\dimen1                        
      \rlap{\hbox to \dimen0{\hfil/\hfil}}      
      #1                                        
   \else                                        
      \rlap{\hbox to \dimen1{\hfil$#1$\hfil}}   
      /                                         
   \fi}                                         %
\catcode`@=11
\long\def\@caption#1[#2]#3{\par\addcontentsline{\csname
  ext@#1\endcsname}{#1}{\protect\numberline{\csname
  the#1\endcsname}{\ignorespaces #2}}\begingroup
    \small
    \@parboxrestore
    \@makecaption{\csname fnum@#1\endcsname}{\ignorespaces #3}\par
  \endgroup}
\catcode`@=12

\usepackage{graphicx}

\begin{document}

\baselineskip=18pt

\setcounter{footnote}{0}
\setcounter{figure}{0}
\setcounter{table}{0}

\begin{titlepage}
\begin{flushright}
HUTP-05/A0051\\
MCTP-05-103
\end{flushright}
\vspace{.3in}
\begin{center}
{\Large \bf  Supersymmetry and the LHC Inverse Problem}

\vspace{0.5cm}

{\bf Nima Arkani-Hamed$^1$, Gordon L. Kane$^2$, Jesse Thaler$^1$, and
Lian-Tao Wang$^1$}

\vspace{.5cm}

{\it $^1$ Jefferson Laboratory of Physics, Harvard University,\\
Cambridge, Massachusetts 02138, USA}

{\it $^2$ Physics Departent, University of Michigan and MCTP \\
Ann Arbor, Michigan 48109, USA}

\end{center}
\vspace{1cm}

\begin{abstract}
\medskip
Given experimental evidence  at
the LHC for physics beyond the standard model, how can we determine
the nature of the underlying theory? We 
initiate an approach to studying the ``inverse map" from the space
of LHC signatures to the parameter space of theoretical models
within the context of low-energy supersymmetry, using 1808 LHC
observables including essentially all those suggested in the
literature and a 15 dimensional parametrization of 
the supersymmetric standard model.  We show that the inverse map of
a point in signature space consists of a number of isolated islands
in parameter space, indicating the existence of
``degeneracies"---qualitatively different models with the same LHC
signatures. The degeneracies have simple physical
characterizations, largely reflecting discrete ambiguities
in electroweak-ino spectrum, accompanied by small adjustments for the
remaining soft parameters. 
The number of degeneracies falls in the range $1<d<100$, depending on
whether or not sleptons are copiously produced 
in cascade decays. This number is large enough to represent a clear
challenge but small enough to encourage looking for new observables
that can further break the degeneracies and determine at the LHC most of the
SUSY physics we care about.  Degeneracies
  occur because signatures are not independent, and our approach
  allows testing of any new signature for its independence. Our
methods can also be 
applied to any other theory of physics beyond the standard model,
allowing one to 
study how model footprints differ in signature space and to test
ways of distinguishing qualitatively different possibilities for new
physics at the LHC.

\end{abstract}

\bigskip
\bigskip


\end{titlepage}

\section{The LHC Inverse Problem}

With the imminent start of the LHC in 2007, particle physics is on
the threshold of its most exciting period in over thirty years. The
secrets of the TeV scale will begin to be revealed, and whatever is
found is likely to have profound
implications for fundamental physics. With only two years to go, what
are the most pressing remaining open
questions in physics beyond the standard model? Over the last two
decades, many models of weak scale physics have been proposed,
starting with the early proposals of supersymmetry \cite{susy} and
technicolor \cite{technicolor}, 
through to the more recent ideas of large and warped dimensions
\cite{ADD, RS}, and
the little Higgs \cite{Arkani-Hamed:2001nc}. The last three years have
seen an explosion of
differing models---ranging from Higgsless models \cite{higgsless},
composite Higgs in
warped compactifications \cite{comphiggs} and other warped
models \cite{warpedgut,warpedgood},  twin Higgs \cite{twinhiggs}, and
SUSY little Higgs models \cite{susylittlehiggs}---largely arising
from combining previous ideas in a
variety of ways. More recently, orthogonal ``un-natural"
directions have also been opened up, inspired by the possibility of
an enormous landscape of vacua in string theory, beginning with the
proposal of split supersymmetry \cite{split} and other minimal models for dark
matter and unification \cite{finetune,finetune-others}.

However, it is clear that the era of speculation and model-building
is near an end---we will very soon get a direct hint for what is
going on from early LHC data. It is unlikely that
exploring the $(N+2)$nd variation on the $(N+1)$st mechanism for
electroweak symmetry breaking is particularly important at this
moment in time.   Unless some qualitatively new ideas and associated
signals are involved, one might as well wait and see
what nature tells us before investing a lot of time on detailed
model variations.

Instead, there is another problem, far more urgent especially as the
LHC draws near, that has received less attention than
model-building. How will we determine the underlying
new physics from LHC data? Of course the first important question
is:  how will we know there is {\it any} new physics beyond the
standard model? It is possible that the signals will be sufficiently
spectacular as to immediately tell us there is new physics and
narrow the space of possibilities to a very
small number. For instance, if the LHC finds evidence for a colored
particle decaying an observable distance from the beamline, this
would look a lot like split SUSY or perhaps other models with nearly
stable colored particles, but not anything like a garden-variety
supersymmetric model.

Even without such dramatic signatures, if there are new colored
particles beneath the $\sim 2$ TeV scale that are produced and
decay to SM particles, we can fairly quickly be convinced of the
existence of new physics. But determining the properties of the
underlying model becomes more challenging, even at a rough level.
For instance, can we know if we have discovered SUSY, or are the
signals with trileptons and missing energy due to extra dimensions
with KK parity \cite{Appelquist:2000nn}, or little Higgs theories with
$T$-parity \cite{Cheng:2003ju}? Despite some recent studies
\cite{Cheng:2002ab,Barr:2004ze,Goto:2004cp,Smillie:2005ar,Cheng:2005as,Datta:2005vx,Datta:2005zs,Barr:2005dz}, not much
systematic work has been done on this ``inverse"
problem \cite{Binetruy:2003cy,Kane:2005az}. Let's  suppose even that we
are working in the context of low-energy SUSY 
with minimal field content. Will we be able to determine even
qualitative properties of the spectrum? Can we, for instance, tell
even roughly whether or not the gaugino masses are consistent with
GUT scale unification? Whether the LSP is a good candidate for Dark
Matter?

\begin{figure}
\begin{center}
\includegraphics[scale=0.5]{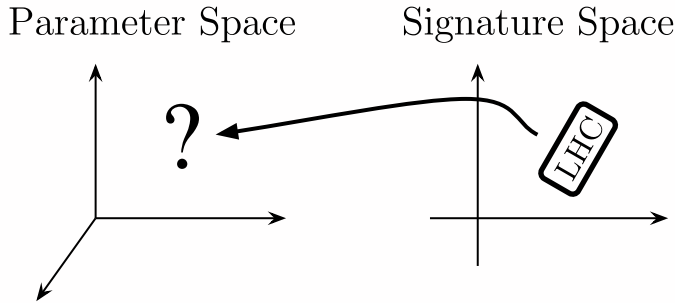}
\end{center}
\vspace{-.1in}
\caption{The Inverse Map from LHC Observables to Theoretical Models.  Given observed signals for physics beyond the standard model, how can we determine the underlying theoretical model?}
\label{fig:inverseintro}
\end{figure}

Instead of addressing this question, most of the work on collider phenomenology to date has been done in
the ``forward" direction, studying the map from parameter
or model space into the space of observable collider
signatures. The signals for a specific model are studied in great detail, with
the goal of seeing how well the parameters of the model can be
measured or constrained. Often, many of the signals are tailor-made to the model at hand and aren't
effective for other models, particularly not for the general case.  To
make the studies tractable, they are usually performed within
simplified models with very few parameters---in the context of SUSY,
for instance, 
these have been carried out with mSUGRA, gauge mediated
and anomaly mediated SUSY breaking
\cite{Baer:1995nq, Baer:1995va,Mrenna:1995ax,gmsb,Hinchliffe:1996iu,amsb,atlastdr},
for several recent studies, see
\cite{Gjelsten:2004ki,Gjelsten:2005aw,Gjelsten:2005vv,Lester:2005je,Lafaye:2004cn,Bechtle:2005qj,Weiglein:2004hn}.  
Presumably, the hope is that if enough
models are simulated in the forward direction, we will gain
familiarity with the associated signals and will be able to spot
them if they arise at the LHC.

\begin{figure}
\begin{center}
\includegraphics[scale=0.35]{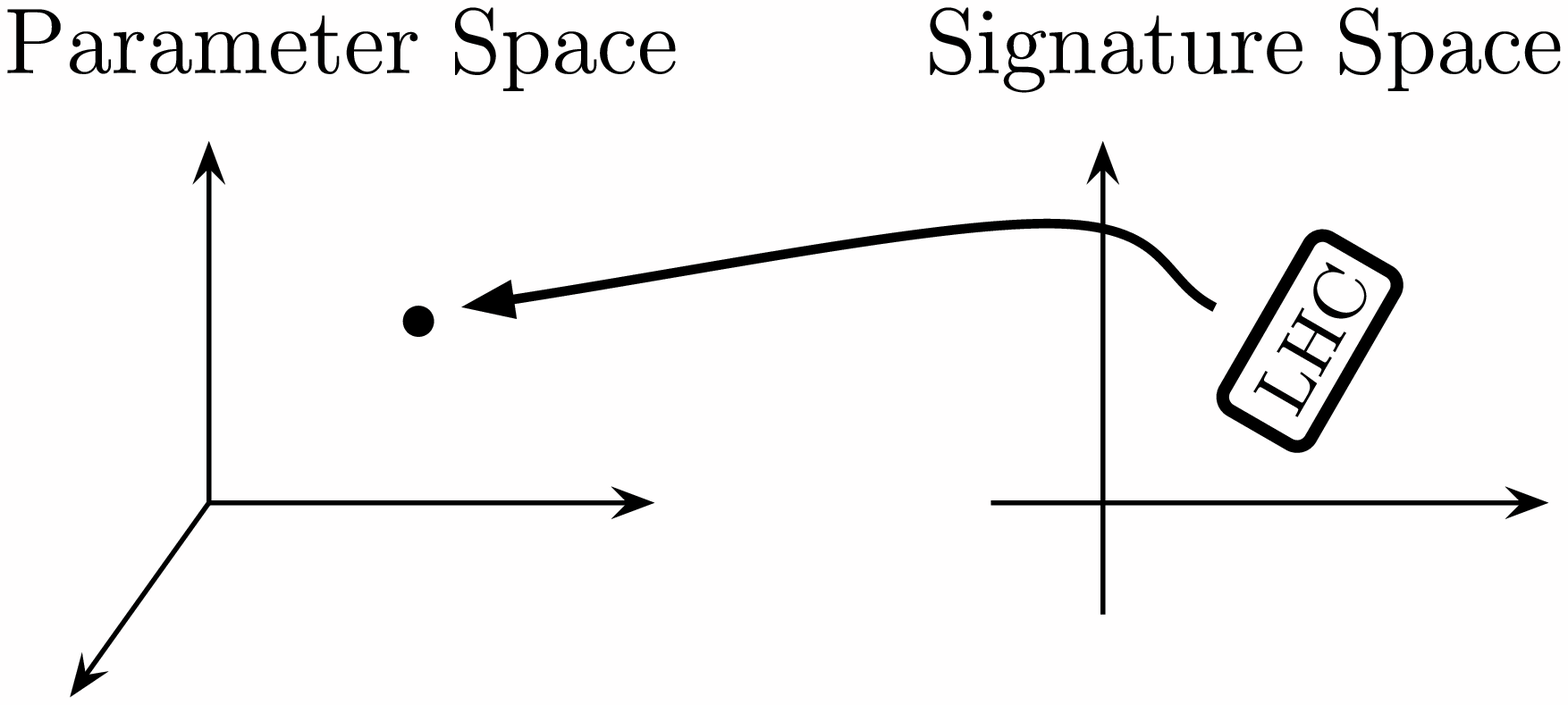} $\qquad \qquad$ \includegraphics[scale=0.35]{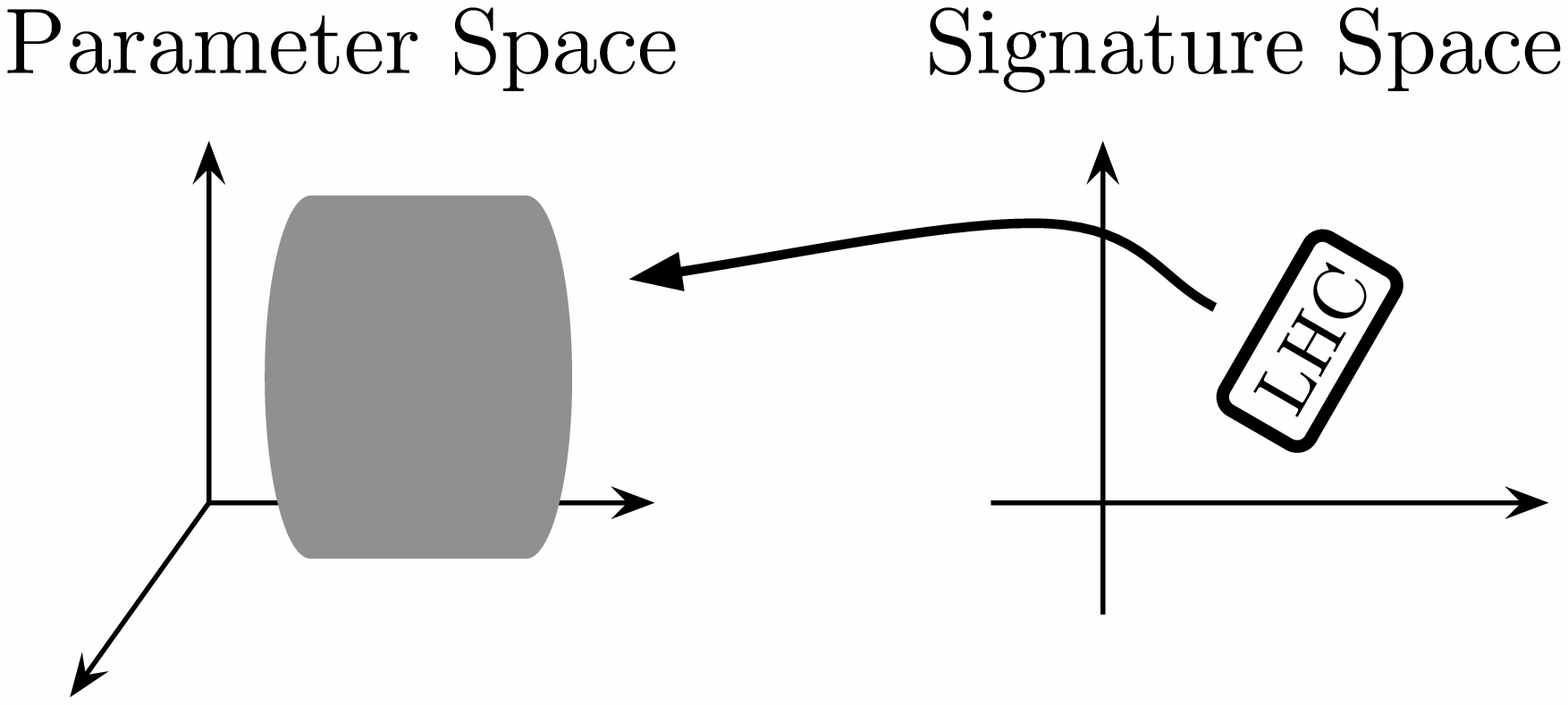}
\end{center}
\vspace{-.1in}
\caption{The Inverse Map in the Best and Worst of All Possible Worlds.
  Ideally, the image of LHC data onto parameter space would specify an
  unique underlying model.  In the most pessimistic scenario, LHC data
  would suggest the physics beyond the standard model without giving
  us any clues as which model describes the new physics.} 
\label{fig:bestworst}
\end{figure}

But the LHC inverse problem---studying the map from LHC signatures to weak scale models as in figure \ref{fig:inverseintro}---is more interesting, important and challenging.  At a hadron collider, it is difficult to directly measure masses and other properties of new particles, a problem exacerbated in SUSY by the escaping LSPs carrying away missing energy. Plus, any signal is likely to receive contributions from multiple channels, and the signals that are reliably observable are more limited. 
A given set of observations at the LHC select a small region point
in signature space, whose size is determined by both by intrinsic
statistical/quantum (``$\sqrt{N}$") fluctuations as well as by
experimental errors. The question is, what does the inverse map back
to parameter space look like? Even very basic issues are open---for
instance, is this map one-to-one?

As shown in figure \ref{fig:bestworst}, in the best of all possible worlds, the inverse map from a small
region in signature space would pick out a small region in parameter
space, so that the underlying model would be uniquely picked out, and better
measurements of the signals would yield a more accurate
determination of the model parameters. In the worst of all possible
worlds, the inverse map into parameter space would fill out a huge
continuously connected region of parameter space, giving us no handle
on even the basic structure of the underlying theory. 

\begin{figure}
\begin{center}
\includegraphics[scale=0.5]{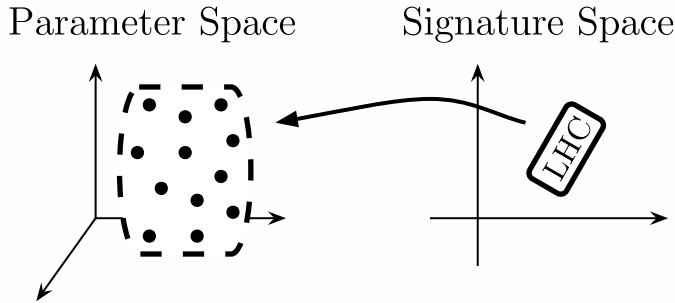}
\end{center}
\vspace{-.1in}
\caption{Our Picture of the Inverse Map.   In the context of
  low-energy supersymmetry, we find that the inverse map of LHC data
  consists of a number of disconnected and widely separated regions in
  parameter space.  This indicates the presence of
  degeneracies---different underlying models that share the same LHC
  signatures.} 
\label{fig:ourinverse}
\end{figure}

In this paper, we initiate a systematic approach to the LHC inverse
problem in the context of low-energy SUSY. As we will see, in SUSY
the actual picture is an intermediate one---the inverse map consists
of a number of isolated islands in parameter space as in
figure~\ref{fig:ourinverse}.   While each island is small, there are a number of them corresponding to qualitatively different models. For obvious reasons we call these
``degeneracies"---different models with the same LHC signatures.
We will be able to give a simple physical interpretations
of the main kinds of degeneracies, which arise largely from discrete
ambiguities in the electroweak-ino sector.

In the next section, we describe our approach to understanding the
map from signature space to parameter space, and describe the simple
statistical techniques that allow us to indirectly infer the
expected number of models with degenerate LHC signals. In section
\ref{sec:ourpicture}, we then summarize our picture of the inverse map, together
with an outline of a  simple physical characterization of the
dominant source of degeneracies. In section \ref{sec:details}, we present the
details of our specific study, including the MSSM parameter space
we scanned and the 1808 LHC signals we considered. In sections \ref{sec:character} and \ref{sec:really}, we give a
more detailed characterization of degeneracies and provide specific
examples of pairs of degenerate models exemplifying each
class of degeneracies.  Part of the reason for the degeneracies is
that, in most of our parameter scan, on-shell sleptons are not
copiously produced in cascade decay chains, so there is less
leptonic information. In section \ref{sec:sleptons}, we do a dedicated scan over a
parameter space where sleptons are typically present in the decay
chains, and find that as expected, the number of degeneracies is
significantly smaller. In sections \ref{sec:future} and \ref{sec:discovery}, we describe possible
improvements and extensions to our work and outline how one should use our study after a real discovery at the LHC. We end with a discussion
and outlook on future work in this area.

\section{From Parameters to Signatures and Back}
\label{sec:bigpicture}

\subsection{Parameter and Signature Space}

Before studying the inverse map from signature space to parameter
space, we should specify what each of these spaces are. Of
course, the most pressing question at the LHC will be to figure out whether there is \emph{any} evidence for physics beyond the standard model, and then most broadly what theoretical framework best describes the new physics---for
instance is it SUSY, or strong dynamics, or something else? As has
been increasingly appreciated over the last couple of years, even
here there are possibilities for degeneracies. In particular, any
model with a stable particle protected by a parity symmetry has
similar looking missing energy signals as SUSY, and it is a
challenge to decisively distinguish the models.  As far as we can
see, there is no systematic approach to this problem yet.

But there is also an analog of the inverse problem purely within the
context of low-energy SUSY. The parameter space of the model is so
huge, and gives rise to such a large range of possible signals, that
we can already ask, even assuming we have low-energy SUSY, whether
we can even roughly determine the correct region of parameter space from LHC data. In this context, the ``parameter space" is clear---fixing for
instance the minimal field content of the supersymmetric standard
model, we can vary the $ \sim 105$ soft parameters of the theory. In
practice, this is much too large a number, and it is useful to
consider a smaller subset that still captures much of the variety of
physics to be expected. Many of the 105 soft parameters are relevant
only to flavor physics and do not have much effect on collider
physics. These consist of three gaugino masses 
$M_{1,2,3}$, the Higgsino mass parameter $\mu$, degenerate soft
masses for the first two generations of $Q, U^c, D^c, L, E^c$ fields (in order to avoid large flavor violations), and separate soft terms for the
third generation scalars. If we also include $\tan \beta$, the ratio of the up- and down-type Higgs vevs, this gives a total of 15 parameters.

The ``signature space" is also easily defined, as we will do in more
detail in section \ref{sec:observables}. After
imposing the appropriate cuts, we associate a number with every LHC observable. For instance, any number count of
events of a particular kind is a direction in signature space.  Any
kinematic histogram can be 
divided up into deciles, and the boundaries of each decile is
a direction in signature space. In this way, we quickly get a huge
dimensionality for signature space---the LHC observables  in our study span
1808 dimensions.

Since signature space has a much higher dimensionality than
parameter space, naively one would think that the inverse map from
signature to parameter space would be unique. Actually, this is
incorrect, largely because the signatures tend to be highly
correlated with each other, so the effective dimensionality of
signature space is much smaller. In fact, we can imagine dividing
signature space up into bins, with size determined by statistical fluctuations and
experimental errors. Even scanning over all possible MSSM
parameters, the number of bins in signature space---the number of
experimentally distinguishable outcomes at the LHC---is not enormously large. One of our results will be a quantitative measure of this
number which will help us understand important aspects of the
inverse map.

An obvious strategy for studying the inverse map is to simply
simulate the MSSM in all regions of its parameter space. For
instance, we can imagine taking each soft mass parameter in 100 GeV
increments between, say, 100 GeV and 1 TeV. Even fixing $\tan \beta$ and
taking our simplified 
$14$ relevant soft parameters at the LHC, this is a total of $\sim
10^{14}$ models. In practice it is impossible to simulate so many
models---simulating the first year of LHC data for a single SUSY model
with fast 
detector simulation takes about 1 CPU hour. This is why so many
studies are performed in models with a much smaller dimension
parameter space, like the 5 dimensional parameter space of mSUGRA.

\subsection{Statistics and the Birthday Problem}

However, we can get a good idea of what the
inverse map looks like even with a large parameter space without
having to simulate so many models. We do this by simulating $m$
models, but then comparing the signatures between all $m (m - 1)/2$
pairs of models. This gives us statistical leverage and also allows
us to determine gross properties of signature space, including
its real effective dimensionality as well as the the total number
of ``bins" or experimentally distinguishable models.

\begin{figure}
\begin{center}
\includegraphics[scale=0.55]{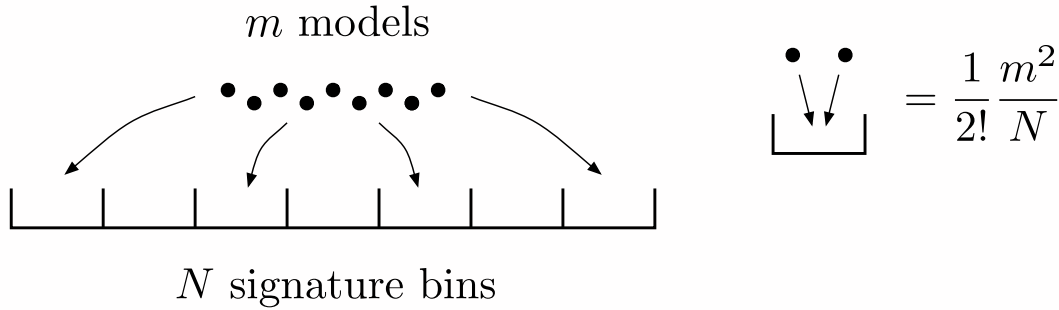}
\end{center}
\caption{The Birthday Problem for the MSSM.  We simulate $m$ models
  and associate each model with a bin in signature space.  For $m \ll
  N$ it is unlikely for any given pair of models to share the same LHC
  signatures, but there is a statistical expectation value $N_2 \sim
  m^2/(2N)$ for the total number of pairs that end up in the same
  bin.} 
\label{fig:birthday}
\end{figure}

We can do all of this because of the famous ``birthday problem'' shown in figure \ref{fig:birthday}. To take a
simple analogy---suppose we throw balls into a box with $N$ bins; we
can't see inside the box so we don't know how big $N$ is. But the balls
are sticky, so that if two balls land in the same bin, they stick
together. We can throw in as many balls as we want, and then empty
the box. One might think that to determine $N$, one has to throw in
$m = N$ balls to cover all possible bins, and then the pigeonhole principle would guarantee seeing a pair of balls stuck together when $m = N+1$, but this is not the case. If we
assume that the balls fall into the bins randomly, on average the
number of bins with $p$ balls $N_p$ is\footnote{This definition of $N_p$ is still approximately true even for large $m$ and small $N$ if we say that a bin with $q$ balls makes a contribution of $q$ choose $p$ to the expectation value of $N_p$.}
\begin{equation}
N_p \sim \frac{m^p}{p! \, N^{p-1}}.
\end{equation}

In particular, the number of doubles is $N_2 = m^2/(2 N)$. So, when
the balls are dumped out of the box, we can see how many doubles
$N_2$ there are, and this allows us to determine $N$ as $N = m^2/(2
N_2)$. Furthermore, if we saw some triples and quadruples too, we
could test our hypothesis by seeing whether the value of $N$ we
extract from all of them are consistent with each other. Clearly,
for this pair-wise counting strategy to be effective, we have to
have $m \sim \sqrt{N}$ to have at least a few doubles, but this is a
big improvement over the naive expectation that we need $m \sim N$.

\begin{figure}
\begin{center}
\includegraphics[scale=0.5]{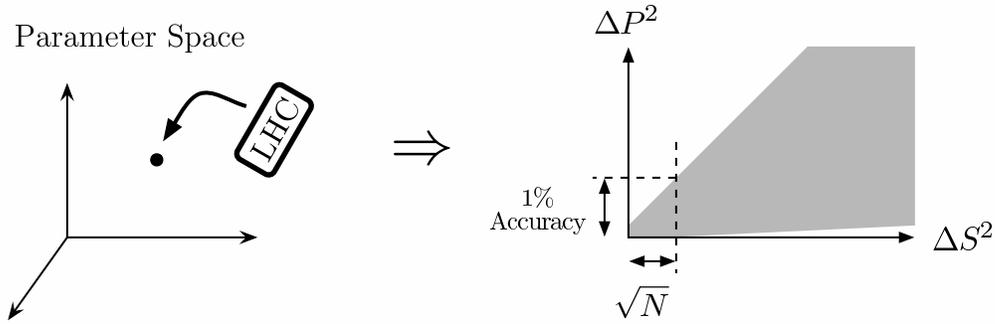}
\end{center}
\vspace{-.1in}
\caption{The $(\Delta S^2, \Delta P^2)$ plot in the best of all
  possible worlds.  The expected confidence range on model parameters
  is defined by the maximum value of $\Delta P^2$ at the $\Delta S^2$
  corresponding to statistical fluctuations.} 
\label{fig:vs-best}
\end{figure}

We follow this strategy in our study. We simulate $m$ models and
associate each with a point in LHC signature space.  Any {\it pair} of
models has an 
associated distance $\Delta P^2$ in parameter space and $\Delta S^2$ in
signature space\footnote{See section~\ref{sec:comparing} for precise
  definitions of $\Delta P^2$ and $\Delta S^2$.}. We can look at a 2D
plot of the points $(\Delta S^2, 
\Delta P^2)$ for all $m(m-1)/2$ pairs of models.  In the
best possible situation, this plot would look like a narrow
triangle, as in figure \ref{fig:vs-best}. The minimal $\Delta S^2$ corresponding to
$\sqrt{N}$ error would correspond to a small $\Delta P^2$ in
parameter space and larger $\Delta S^2$ would be associated with
larger $\Delta P^2$.

\subsection{Degeneracies}
\label{sec:degens}

\begin{figure}
\begin{center}
\includegraphics[scale=0.5]{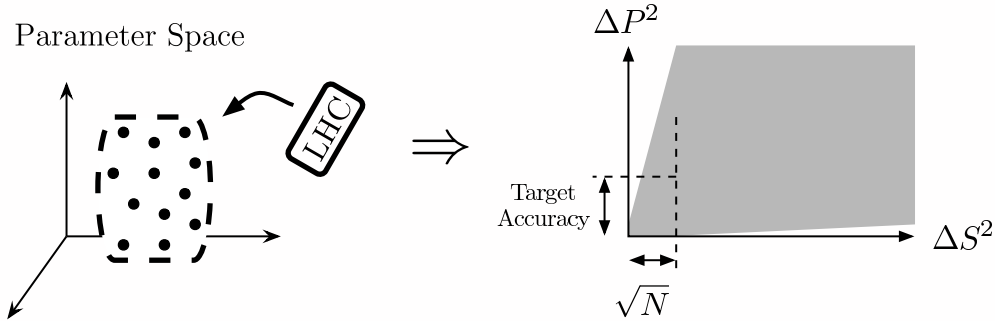}
\end{center}
\vspace{-.1in}
\caption{A cartoon of $(\Delta S^2, \Delta P^2)$ plot in the general
  MSSM.  At the $\Delta S^2$ value corresponding to statistical
  fluctuations, the maximum value of $\Delta P^2$ exceeds the target
  accuracy on model parameters.   While it is possible to change the
  target accuracy value to formally decrease the number of
  degeneracies, for any reasonable choice (such as 10\% or 20\%
  error), there is still a sizable number of degenerate pairs.
  Notice that target accuracies of the parameters usually correspond
  to a much larger region in the parameter space than that is occupied
  any small island. This corresponds to the existence of many ``cliffs'',
  see section~\ref{sec:cliffs}.}
\label{fig:vs-ours}
\end{figure}

\begin{figure}
\begin{center}
\includegraphics[scale=0.9]{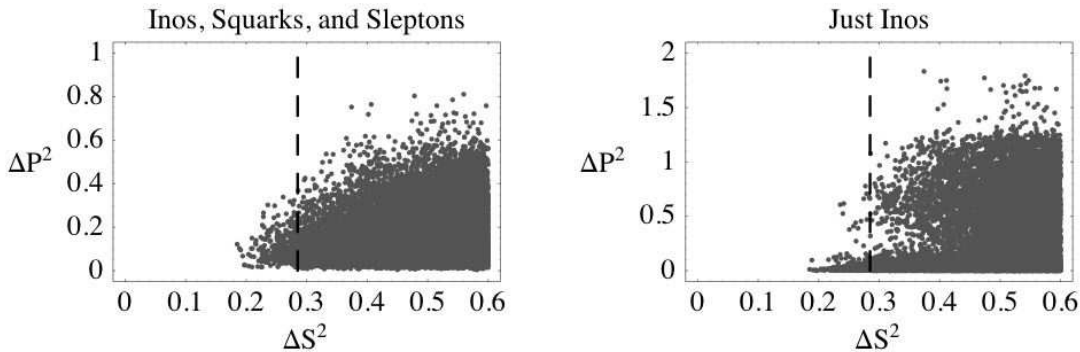}
\end{center}
\caption{The $(\Delta S^2, \Delta P^2)$ plot for the MSSM.   In the
  left plot, we measure $\Delta P^2$ using the 14 mass parameter we
  scan, where $\Delta P^2 \sim 0.01$ roughly corresponds to 10\%
  accuracy in all parameters.  The position of the vertical dashed
  line corresponds to 
  the distance in signature space from statistical fluctuations, so
  the fact that there is a sizable number of model pairs to the left
  of this line with large $\Delta P^2$ indicates the presence of
  degeneracies.  The right plot shows the same models with $\Delta P^2$
  measured only using the gaugino and higgsino mass parameters.  As we
  will explain in section \ref{sec:ourpicture}, the break in the plot
  between small and large $\Delta P^2$ comes from a bimodal behavior
  in the electroweak-ino sector, which explains the dominant reason
  why there are degeneracies in the MSSM.  Note that only a small
  number of pairs are shown on these plots; the $\Delta S^2$ values
  extend out to $\Delta S^2 \gsim 600$.} 
\label{fig:svsp-master}
\end{figure}

The real situation with the MSSM is rather different---the plot for
the models we simulate is shown in figures \ref{fig:vs-ours} and \ref{fig:svsp-master}. We will specify our
parameters and signatures in more detail in section \ref{sec:details}. For
the present it suffices to say that $m = 43026$ MSSMs were
simulated. The vertical dashed line in figure \ref{fig:svsp-master}
marks $\Delta S^2 = 0.285$, corresponding to the distance in signature
space when the {\it same model} is repeatedly simulated, giving us a measure of the size of statistical fluctuations. 

The most striking feature of this plot is the clear presence of a
relatively large number of degenerate models---there are many points
with small $\Delta S^2$ but large $\Delta P^2$.  (As we explain in section \ref{sec:comparing}, $\Delta P^2 \sim 0.01$ corresponds to parameters matching to 10\% accuracy.)  The $43026$ models
represent a very sparse sampling of the 14 dimensional parameter
space, and yet $283$ pairs of models share indistinguishable
signatures!

From our ``birthday problem" picture, this is clear evidence that
the number of effective ``bins" in signature space is not enormously
large. The expected number of signature bins is
\begin{equation}
N_{\rm sig} \sim \frac{m^2}{2 N_2} \sim 3.3 \times10^6.
\end{equation}
As described
in appendix \ref{sec:demonic}, we can confirm that this picture is correct by extracting a value of
$N$ also by looking at the number of triplets and quadruplets of
models with the same signatures. The value of $N$ extracted in this
way is indeed of the same order of magnitude as above.  The fit can be further improved by
taking into account that some fraction of the bins are more likely
to be populated than others.

\begin{figure}
\begin{center}
\includegraphics[scale=0.6]{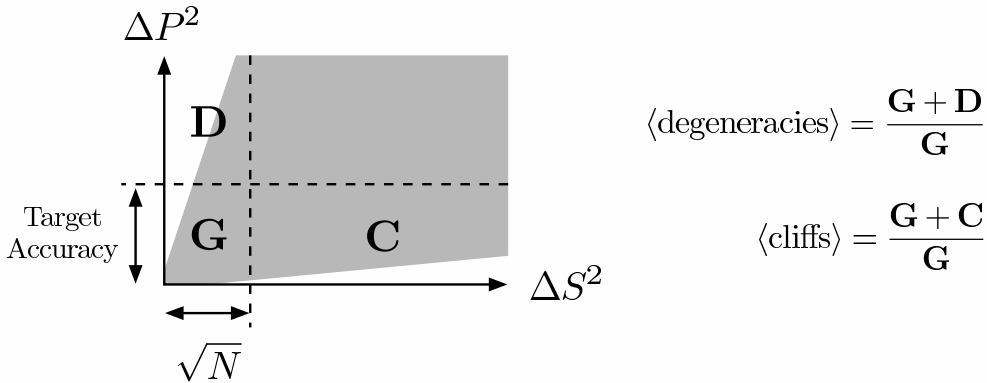}
\end{center}
\vspace{-.1in}
\caption{A cartoon showing how to count the expectation value for the number of degeneracies and cliffs.  Region $\mathbf{G}$ corresponds to ``good pairs'' of models that are close both in parameter space and signature space.  Region $\mathbf{D}$ correspond to pairs of models that are degenerate, \emph{i.e.}  close in signature space but well separated in parameter space.  Region $\mathbf{C}$ corresponds to ``cliffs'', where the distance between models in signature space is large despite the models' proximity in parameter space.}
\label{fig:degen-cartoon}
\end{figure}

Our estimate for $N_{\rm sig}$ gives us a good idea of the number of
distinguishable bins in signature space---that is, the number of {\it
possible distinguishable} MSSMs in the parameter space we scanned.
But the actual number of degeneracies we care about is criterion
dependent. For instance, if two models are degenerate with mass
parameters differing by, say, only 5\% from one another, then we may
not wish to count these models as being ``really different".  We can define what
we mean by ``good pairs" which are sufficiently close in parameter space that we would expect them to be close in signature space.  Given
such a criterion, the expectation value for the number of degeneracies from figure
\ref{fig:degen-cartoon} is
\begin{equation}
\label{eq:degen}
\langle \mbox{degeneracies} \rangle = \frac{\mbox{number of pairs close in signature space}}{\mbox{number of ``good" pairs}}.
\end{equation}

\begin{figure}
\begin{center}
\includegraphics[scale=0.9]{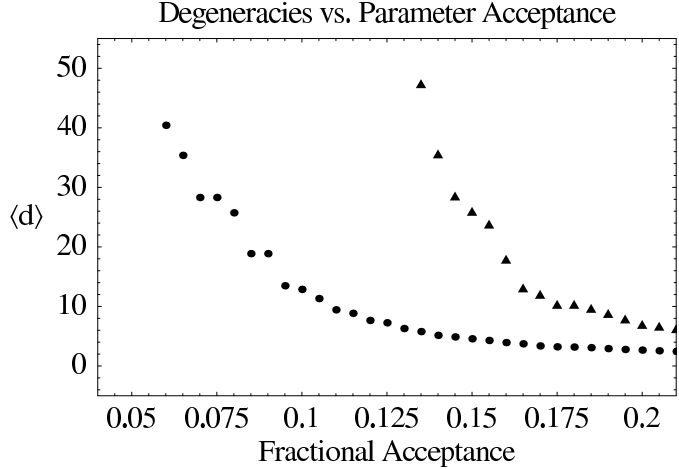}
\end{center}
\caption{Number of degeneracies as a function of allowed fractional
  error.  Triangles correspond to imposing fractional acceptance on
  gaugino, higgsino, and squark parameters.  Dots correspond to  just
  imposing fractional acceptance on the inos.  The fact that the number
  of degeneracies asymptotes to 2 indicates that the
  degeneracies are not caused simply by the presence of large error
  bars on model parameters but also because of discrete choices in the
  spectrum.} 
\label{fig:degen-percent}
\end{figure}

For instance, for our set of 283 LHC indistinguishable pairs, we can
decide to declare two models ``the same" if they have the same
ordering of the ino spectrum.  In this case, the number of
degeneracies is $\langle d \rangle = 4.4$---that is, for a given model, there are about 4 other models with the same LHC
signatures but a {\it different} ino ordering.   Or we can decide that
two models with ino masses within 10 percent of each other are
``the same".  In this case, $\langle d \rangle = 12.9$, so we expect for
any given model, there are about 12 others with the same LHC
signatures but with ino masses differing by more than 10
percent.  There are only 2 out of the 283 pairs whose ino and squark
masses match to 10\%, so using that criteria we would expect $\langle
d \rangle \sim 140$.  In figure \ref{fig:degen-percent}, we show how
the number of degeneracies changes as we adjust the fractional acceptance
in parameter space between pairs of models, showing that the source of
degeneracies is not coming only from large error bars on individual
parameters but also from discrete choices in the SUSY spectrum. 

So we see that the number of degeneracies is ${\cal O}(10 - 100)$
with the criteria we have specified. We can get a rough idea of
whether this number is in the right ballpark by estimating the
number of degeneracies in a different way. We have determined that
there are $N_{\rm sig} \sim 3 \times 10^6$ possibly distinguishable
MSSMs. We can also estimate the number of ``different" models
$N_{\rm models}$ we have in our parameter space based on the parameter ranges given in section \ref{sec:masses}.  Assuming that LHC signatures are sensitive to gluino mass variations of $50 \GeV$, squark mass variations of $75 \GeV$, and electroweak-ino mass variations of $100 \GeV$, then\footnote{These values are chosen by estimating the local variation in mass parameters from the $(\Delta S^2, \Delta P^2)$ plots in section \ref{sec:character}.}
\be
N_{\rm models} \sim 8 \times 5^6 \times 9^3 \sim 10^8.
\ee
Since $N_{\rm models} > N_{\rm sig}$, by the pigeonhole principle there {\it
must} be degeneracies.  Furthermore, a rough estimate for the number
should be
\begin{equation}
\label{eq:degenalt}
\langle \mbox{degeneracies} \rangle \sim \frac{N_{\rm models}}{N_{\rm sig}}
\sim 30
\end{equation}
which is indeed consistent with our first estimate using equation (\ref{eq:degen}).

\subsection{Cliffs}
\label{sec:cliffs}
Another important feature of the $(\Delta S^2, \Delta P^2)$ plot is the
population of points along the horizontal axis, which shows that
there are models with small distance in parameter space but large
differences in signature space.  These indicate the existence of
``cliffs" in model space---small parameter changes can give rise to
large changes in the signatures. In particular, this suggests that in
any local region of parameter space, the map to LHC signature is
essentially one-to-one.

We can easily quantify this notion. Given any criterion for whether
two models share the same ``parameter bin'', we can find the total number of pairs of models which
are close in parameter space and compare with the number which are
close in both signature space and parameter space.  This defines an expected value of cliffs, \emph{i.e.} the
chance that two models within the same parameter bin end up in the same signature bin.
\begin{equation}
\langle \mbox{cliffs} \rangle = \frac{\mbox{number of pairs close in parameter space}}{\mbox{number of ``good'' pairs}}.
\end{equation}
We can look at a plot of the number of cliffs as a function of
changing the signature cut that defines ``good'' pairs.   Depending on
the criteria for closeness in parameter space, we see that the number
of cliffs is $\mathcal{O}(10^3)$--$\mathcal{O}(10^4)$, showing that a single bin in
parameter space maps to many different bins in signature space.  This
confirms the picture in figure \ref{fig:ourinverse} that the isolated
regions in parameter space that map to the same signature bin are
indeed ``small''. 

\begin{figure}
\begin{center}
\includegraphics[scale=0.9]{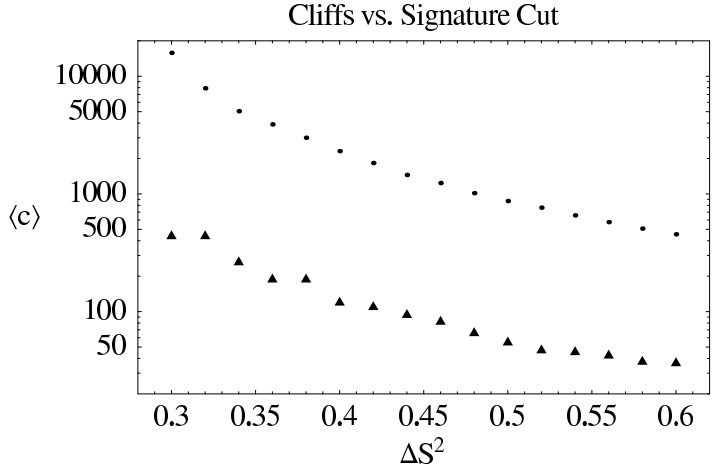}
\end{center}
\caption{Number of cliffs as a function of signature cut.  $(\Delta S)^2 = 0.285$ corresponds to statistical fluctuations.  Triangles correspond to imposing 10\% percent acceptance on gaugino, higgsino, and squark parameters.  Dots correspond to imposing 10\% percent acceptance on just the inos.  The large number of cliffs indicate the strong sensitivity of LHC observables in local regions of parameter space.}
\end{figure}

\subsection{Effective Dimensionality of Signature Space}
\label{sec:dimsig}

Formally, our signature space is very high dimensional.    Of course, we are mapping a 15
dimensional parameter space onto signature space, so we do not expect the signature space
manifold spanned by SUSY models to be more than 15 dimensional.  However,
the existence of degeneracies strongly suggests that there are huge
correlations between the signatures, such that the actual
dimensionality of the space populated by SUSY models is far smaller.

\begin{figure}
\begin{center}
\includegraphics[scale=0.9]{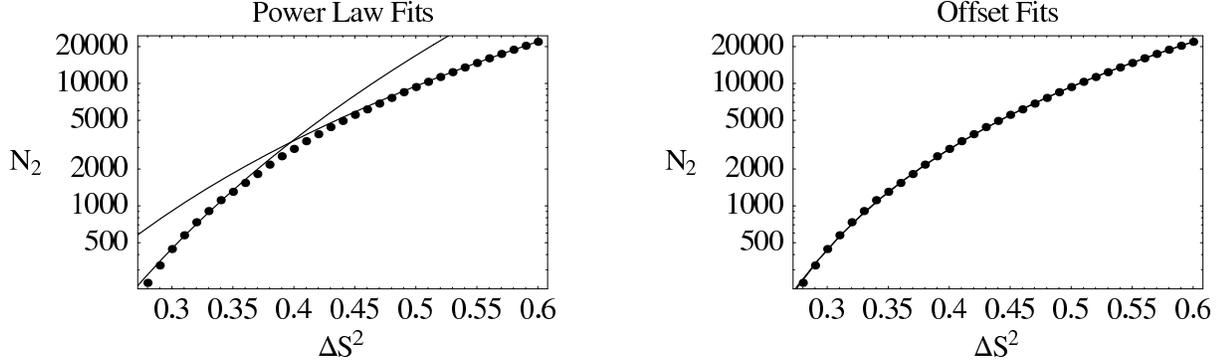}
\end{center}
\caption{Dimensionality of signature space.  $(\Delta S)^2 = 0.285$ corresponds to statistical fluctuations, and the maximum value of $(\Delta S)^2$ between pairs of models is $\Delta S^2 \gsim 600$, well off these plots.  On the left plot, we fit to the power law $N_2 \sim (\Delta S)^{D_{\rm sig}}$ with exponents $D_{\rm sig} \sim 9 \mbox{ and } 14$.  On the right plot, we take into account the possibility of an offset value $\Delta S_0$, yielding $D_{\rm sig} \sim 5 \mbox{ or } 6$ (both curves are shown overlapping).  In either case, we see that the effective dimensionality of signature space is much smaller than the dimensionality of parameter space, giving another justification for degeneracies.}
\label{fig:dimsigspace}
\end{figure}

We can quantify this simply by looking at
how the number of degenerate model pairs $N_2$ depends on the
distance in signature space $\Delta S$; if signature space is
effectively $D_{\rm sig}$ dimensional we expect
\begin{equation}
N_2 \sim (\Delta S)^{D_{\rm sig}}.
\end{equation}
Now, the precise definition of $\Delta S$ requires some care.  If we take the definition
from section \ref{sec:comparing} and fit to $N_2$, we find in figure \ref{fig:dimsigspace} that
\be
N_2 \sim \left\{   \begin{array}{ll}(\Delta S)^{9.2}  & \mbox{near $(\Delta S)^2 = 0.3$} \\
(\Delta S)^{14.2} & \mbox{near $(\Delta S)^2 = 0.6$} \end{array} \right. ,
\ee
suggesting that the dimensionality of signature space in the vicinity of degenerate pairs
is 9 to 14 dimensional.

Still, the fact that $N_2$ does not seem to follow a simple power law in $\Delta S$ indicates that
$\Delta S$ may not the best measure of distances in signature space.  In particular, even simulating
the same model repeatedly gives some finite value of $\Delta S$, so
one might guess that for a signature space of dimension $D_{\rm sig}$:
\begin{equation}
N_2 \sim (\Delta S - \Delta S_0)^{D_{\rm sig}} \quad \mbox{or} \quad N_2 \sim
\left(\sqrt{\Delta S^2 - \Delta S_0^2}\right)^{D_{\rm sig}}.
\end{equation}
In figure \ref{fig:dimsigspace}, we see that this hypothesis is valid over a wide range of
signature values, simultaneously giving good fits to\footnote{The number $\Delta S_0 = 0.42$ or $0.39$ is interesting, because it is
around the minimum $\Delta S$ between identical models run with different random number seeds.}
\begin{equation}
N_2 \sim \left(\sqrt{(\Delta S)^2 - (0.42)^2}\right)^{6.2} \quad \mbox{and} \quad N_2 \sim
\left(\Delta S - 0.39\right)^{4.4}.
\end{equation}

These fits suggest that the true dimensionality of signature space is $D_{\rm sig} \sim 5$ or
$6$, less than the number of MSSM parameters that we are varying. This makes it clear that
in order to further break degeneracies, it is
not enough to add a few extra signatures; one must add signals that
are sufficiently orthogonal to the existing ones to increase the
effective dimensionality of signature space.

Ideally, we could figure out what the $5$ or $6$ independent signatures are in a statistical sense by doing a multivariate regression on our LHC signatures.  Unfortunately, it is computational difficult to do such a regression on a 1808 dimensional signature space.  Moreover, we expect at least some of the independent signature directions to change depending on the specifics of the MSSM spectrum.

\section{Our Picture of the Inverse Map}
\label{sec:ourpicture}

The existence of degeneracies and cliffs substantiates the rough
picture of the inverse map we suggested in figure \ref{fig:ourinverse} consisting of a number of
small islands spread out over a large region in parameter space. The
small size of the islands is reflected in the existence of
cliffs, indicating that LHC observables are indeed sensitive to small parameter changes.
The existence of many islands far apart in parameter space is substantiated by the presence of degeneracies.

Note that the figure assumes that the islands are ``point-like" in
parameter space.  We actually don't know that this is the case---they
may be higher-dimensional manifold like tubes or sheets in parameter
space.  Indeed, if there are obvious flat directions in parameter
space where for instance a particle is sufficiently decoupled as to
play little role in the signatures, these would show up as
higher-dimensional objects in the inverse map. All we can say with
confidence from our analysis of cliffs is that the
islands are not space-filling and constitute a small part of
the total volume of the inverse map from  a given signature.

\begin{figure}
\begin{center}
\includegraphics[scale=0.7]{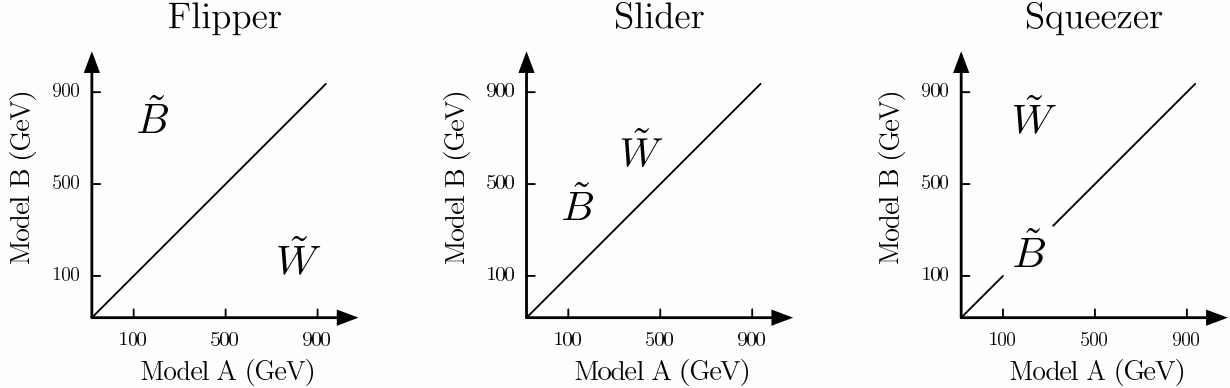}
\end{center}
\caption{Dominant Degeneracies:  Flippers, Sliders, and Squeezers.
  When scanning over the general MSSM, sleptons are generically
  decoupled, therefore important leptonic handles on the
  electroweak-ino spectrum are lost.   This allows for three major
  classes of degeneracies.  In a ``flipper'', the identity of an
  electroweak-ino is changed while the mass eigenstates are fixed.  In
  a ``slider'', the entire spectrum is shifted while fixing mass
  differences.  In a ``squeezer'', a low-energy eigenstate is
  hidden because the decay products between adjacent mass eigenstates
  are soft.  In the above cartoons, ``$\tilde{B}$'' and
  ``$\tilde{W}$'' can stand for a bino, a wino, or a higgsino.}   
\label{fig:flipper-cartoon}
\end{figure}

The degeneracies have simple physical interpretations as shown in figure \ref{fig:flipper-cartoon}. As is common
in hadron collider phenomenology, the cleanest handle on new physics
often comes from looking at leptons. In SUSY models where sleptons
are not copiously produced in a long SUSY cascade decay chain, the
leptons dominantly come from $W$s and $Z$s produced in electroweak-ino cascade
decays. In this case, we find large ambiguities in the spectrum of
the remaining electroweak superpartners. Two models can have
identical LHC signals by having ``flippers" where electroweak-ino mass
eigenvalues are fixed but with different eigenstates, ``sliders"
where the electroweak-ino spectrum is moved up or down keeping mass differences
fixed, and ``squeezers" where the information of some of the electroweak-inos
is hidden because the mass splittings are small enough that the
leptons in the decay products are too soft to be seen.  We will discuss these degeneracy classes further in section \ref{sec:character}.

Of course, the changes in the electroweak-ino sector are accompanied
by suitable changes in the 
colored superparticle spectrum to match rates and other kinematical
distributions between the models. The number of degeneracies for a
given model arising in this way is of order $10$ to $100$.    As we
show in section \ref{sec:sleptons}, when 
sleptons are forced to be present in the cascade decays, there are
more leptons in final states 
and the degeneracies virtually disappear, although there may still be
an ambiguity in swapping 
the left-and right-handed sparticle spectrum. But in the general case,
our estimate of the number 
of degeneracies is as interesting as it could have been---the number
is not one (\emph{i.e.} there certainly {\it are} degeneracies) but nor is
it $10^6$. Therefore while the existence of degeneracies represents
a challenge, it is one that can likely be overcome by devising
clever new observables to eliminate them. This would allow us to
determine essentially {\it all} important aspects of SUSY physics
with LHC data.



\section{Details of Our Study}
\label{sec:details}

Even though the birthday problem has reduced the number of models to
simulate from $m \sim N$ to $m \sim \sqrt{N}$, the number $N$ of
experimentally distinguishable possibilities at the LHC is still quite
large, and there a number of compromises one must make when comparing
a large number of different models.


The first compromise is on the amount of data we can generate.  For
gluino masses around 600 GeV, there are around $10^7$ SUSY events at
an integrated luminosity of 10 fb$^{-1}$.  Even without including the
effect of initial
state radiation and multiple interactions, it takes a fast CPU over
one hour to simulate that many events, and a simplified version of the
event record is roughly 50 megabytes in size.  Most previous collider
studies assume that the LHC will achieve an integrated luminosity of
300 fb$^{-1}$, but it is simply impractical for us to generate, store,
and analyze that much data for a large number of models.  Therefore, we only
generate 10 fb$^{-1}$ of data for each model and force all colored
particles to be heavier that 600 GeV.  While the constraint of heavy
colored particles was necessitated by computational limitations, there
are independent reasons for making this choice.  The characteristic
signature of SUSY at the LHC is hard jets plus missing transverse
energy, so while the cross section for SUSY increases as the gluinos
and squarks get lighter, the jets from SUSY cascades get softer,
making it more difficult to identify a pure sample of SUSY events. 

The second compromise is on standard model background.  While it is
certainly possible to estimate the effect of standard model background
on 10 fb$^{-1}$ of data, the focus of this paper is not on separating
SUSY signals from standard model background but on distinguishing
between different SUSY models.  At 300 fb$^{-1}$, one can make hard
cuts and still maintain decent statistical significance of leptonic
signatures, but this is difficult at 10
fb$^{-1}$.  Therefore, we ignore standard model background in this
study except as a guide for defining reasonable cuts and triggers.

This is simultaneously an optimistic and pessimistic choice, because
by ignoring standard model background we are inflating the statistical
significance of our small data sample, but by ignoring the effect of
higher luminosity we lose access to rare processes that may give
important clues in deciphering the data.  We emphasize that both of
these compromises are not intrinsic limitations, but are dictated by
our current computing resources.

\subsection{LHC Observables}
\label{sec:observables}

Broadly defined, LHC data is anything that can be measured with an
ATLAS- or CMS-like detector with delivered luminosity.  To simulate 10
fb$^{-1}$ of LHC data, we use PYTHIA \cite{pythia} to generate parton level
interactions and hadron showering and pipe PYTHIA output to a modified
version of the CDF fast detector simulator PGS written by John Conway
\cite{pgs}.   This modified version
was developed by Steve Mrenna and approximates an ATLAS- or CMS-like
detector.   PGS yields reasonable efficiencies and fake rates and
includes the effect of energy smearing.

We use a simplified output from PGS for our study, namely a list of
objects in each event labeled by their identity (photon, electron,
muon, hadronic tau, jet, b-tagged jet, missing $E_T$) and their
four-vector. Leptonic objects are also labeled by their charges.
Using this information, one can construct almost any
LHC signature imaginable.    Of course, when real data from the LHC
arrives, various different techniques will be used to isolate, verify,
and make measurements on samples.  Because we are ignoring standard
model background in our analysis, our goal is to choose a set of
observables that are sensitive to MSSM parameters but which are
sufficiently inclusive to be useful over a wide range of parameter
values.  Note that we make no attempt to interpret any of our signatures in terms of cross sections, branching ratios, or mass differences in the underlying model; instead, we simply compare raw signature values between different models.

While we are not including standard model background, initial state
radiation, or multiple interactions in our analysis, in appendix \ref{sec:observelist},  we select
cuts and triggers in a way that is aware of the challenges they pose.
We will focus on events with 2 or more jets plus large missing
transverse momenta, for while jet-veto signatures from direct
production of electroweak-inos can sometimes be important, the standard
model background is generically too large for the parameter region we
scan.

The complete list of the signatures we use is given in appendix
\ref{sec:observelist}.  There are two different types of signatures we
consider, counting signatures and kinematic histograms.  Counting
signatures give the number of events that pass a certain set of
criteria.  Because getting an accurate measurement of $\sigma_{SUSY}$
is very difficult at the LHC, we only include one signature that
counts the total number of SUSY events that pass the above cuts; all
other counting signatures are given as ratios.  The two types of
kinematic histograms we generate are effective mass\footnote{In other
  contexts, ``effective mass'' means the $P_T$ sum over missing energy
  and the four hardest jets.  In this paper, we use effective mass to
  mean any $P_T$ sum over any number of objects.} and invariant mass 
for various different objects in events:
\begin{equation}
m_{\rm eff} = \sum_a P^a_T, \qquad m_{\rm inv}^2 = \left(\sum_a
p^a_\mu \right)^2.
\end{equation}

We use a quantile method to define signatures for all of the
kinematical distributions. The entries in a distribution are organized
into bins of variable width such that each bin contain the same number
of entries. For example, a decile distribution has ten bins which each
contain $10 \%$ of the total entries.   The signatures for a
distribution are given by the boundaries of the bins, with no
signature stored for the upper or lower boundaries of the total
distribution.  Therefore a distribution split into deciles has 9
signatures associated with it.  Note that using the quantile method,
the information of a distribution is in the positions of the bins,
rather than the content of each bin.  

Our adoption of the quantile method is dictated by the necessity of
defining distribution observables which are applicable to a wide
variety of models with very different mass spectra. The traditional
histogram method of fixed bin size is not practical here since we would have to include
histograms with large energy ranges and with many different
bins. Because different parts of the histogram would be populated by
different models, we would either be forced to store a large number of
redundant signatures or make the bin size so large that important
kinematic information could be lost. The quantile method resolves this
problem by converting distributions into a small number of easy-to-use
inclusive observables. 

In the SUSY literature, there is a large focus on endpoints and edges
in kinematic distributions as ways of constraining mass differences
between different SUSY particles \cite{atlastdr,Gjelsten:2004ki,Gjelsten:2005aw,Lester:2005je}.   While it is difficult to teach a
computer how to generically find an edge/endpoint (and assign
appropriate error bars), the quantile method of describing histograms
captures most of the statistically significant information in a
kinematic distribution.  So even though there is not a separate
signature corresponding to an edge/endpoint measurement, the
edge/endpoint will be well constrained by the requirement that all of
the quantiles for the histogram match.

\subsection{The Scanned Parameter Space}
\label{sec:masses}

LHC signatures are mainly sensitive to mass parameters for the
particles which either have large production cross sections or are
important links in cascade decays. Considering both the strong
constraint from FCNC measurements and the lack of sensitivity in LHC
observables included in the current study, we have imposed
universality conditions on the masses of the first two generations of
scalar fermions. On the other hand, third generation mass parameters
are treated as independent. Flavor
off-diagonal entries in the sfermion mass matrices are assumed to
vanish. Left- and right-handed sfermions are allowed to have
independent mass parameters.

We scan $\tan \beta$ and the following 14 SUSY mass parameters:
 \bea 
\mathrm{Inos:} &\ & M_1, \ M_2, \ M_3, \ \mu \nonumber \\
\mathrm{Squarks:} &\ & m_{\tilde{Q}_{1,2}}, \ m_{\tilde{U}_{1,2}}, \
m_{\tilde{D}_{1,2}},\  m_{\tilde{Q}_{3}}, \ m_{\tilde{t}_{R}}, \
m_{\tilde{b}_{R}} \nonumber \\
\mathrm{Sleptons:} &\ & m_{\tilde{L}_{1,2}}, \ m_{\tilde{E}_{1,2}}, \
m_{\tilde{L}_{3}}, \ m_{\tilde{\tau}_{R}}  \nonumber
\eea 
We have separated left- and
  right-handed sfermions. The soft masses for the first two
 generations are universal. For 
  right-handed squarks, the up- and down-types are varied 
  independently.  In order to constrain the Higgs sector, we take the
 mass of the 
pseudoscalar Higgs to be 850 GeV.  We also fix the third generation squark
$A$-terms at 800 GeV.

\begin{figure}
\begin{center}
\includegraphics[scale=0.7]{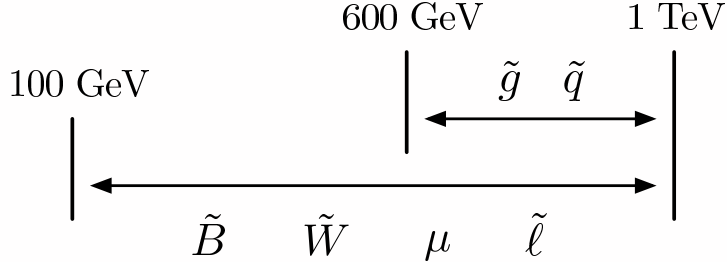}
\end{center}
\caption{Parameter Ranges for our General Scan.  The gluino and the
  six squark masses are randomly sampled between  600 GeV and 1 TeV.
  The three electroweak-ino and the four slepton masses are randomly sampled
  between 100 GeV and 1 TeV.  The value of $\tan \beta$ is scanned
  between 2 and 50.} 
\label{fig:mass-general}
\end{figure}

We scan the parameter space by randomly sampling the 15
dimensional parameter space.  We sample all mass parameters
uniformly in mass value, and scan $\tan \beta$ uniformly from
2 to 50.  The range of mass parameters in our general scan is shown in figure
\ref{fig:mass-general}.

Note that we do not enforce the mass of the Higgs to be above the experimental
bound.  The reason is
that in order to get a heavy enough Higgs in the MSSM, we would be
forced into a small
region of parameter space with large $\tan \beta$ and large stop
masses.  Because the LHC
signatures of SUSY are dominated by colored particles and are largely
independent of the
Higgs sector, we do not want to limit the kinds of allowed SUSY
signatures by placing  artificial restrictions on the SUSY parameters,
especially because the Higgs mass may be  lifted due to an extended
Higgs sector.

We also impose a non-decoupling criterion to minimize the number of flat directions in the map from parameter space to signature space. Let $m_{\rm slepton}^{\rm
  max}$ be the heaviest slepton soft mass, $m_{\rm ino}^{\rm max}$ be
the heaviest electroweak-ino mass parameter ($M_1$, $M_2$, or $\mu$),
and let $m_{\rm color}^{max}$ be the soft mass or mass parameter for
the heaviest color-charged particle.  We demand: \be m_{\rm
slepton}^{\rm max} < m_{\rm ino}^{\rm max} + 50 \GeV < m_{\rm
  color}^{max} + 100 \GeV.
\ee The purpose of the non-decoupling criterion was to ensure that
apart from the $50 \GeV$ buffer, there was always a kinematically
allowed cascade decay involving sleptons in the sample.  However,
this non-decoupling criterion in no way guarantees that sleptons
will play a dominant role in the decay chain. As we discuss more in
  sections \ref{sec:character} and \ref{sec:sleptons}, scanning over
  the MSSM, the total production rate for sleptons from cascade decays
  is generically small.  

\subsection{Comparing Models}
\label{sec:comparing}

The most statistically sound method for testing whether a set of LHC
observables matches a given model is  to generate an ``infinite''
statistics sample of that model, and do a $\chi^2$ test to estimate
the likelihood that the LHC data matches that model.  Because we are
focusing on testing whether there are SUSY models that share
the same signatures with 10 fb$^{-1}$ of data, we only need to find a
measure of the difference between data sets, and then establish a
threshold below which models are considered degenerate.

We define a $\chi^2$-like variable to measure the difference between two models
\be
(\Delta S_{AB})^2 = \frac{1}{n_{\rm sig}} \sum_i \left( \frac{s^A_i -
  s^B_i}{\sigma^{AB}_i} \right)^2, 
\ee
where $s^A_i$ ($s^B_i$) is the value of the $i$-th signature for model
$A$ ($B$), $\sigma^{AB}_i$ is the error bar assigned between models
$A$ and $B$ for the $i$-th signature, and the sum over $i$ runs over $n_{\rm sig}$
relevant signatures.  A relevant signature is one that would not artificially
reduce $\Delta S_{AB}$ because of low statistics.  We define a
relevant signature to be one for which the error bar $\sigma^{AB}_i$
is smaller than both $s^A_i$ and $s^B_i$ or for which $|s^A_i - s^B_i|
> \sigma^{AB}_i$.  The value of $\sigma^{AB}_i$ is given by
\be
\sigma^{AB}_i = \sqrt{\left(\delta_{\rm stat} s^A_i \right)^2  + \left(\delta_{\rm stat} s^B_i \right)^2 + \left(f_i \frac{s^A_i + s^B_i}{2}   \right)^2},
\ee
where the statistical errors $\delta_{\rm stat}$ are described in
Appendix \ref{sec:errors}, and $f_i$ is an additional fractional error
parameter that could be used to estimate standard model background
errors.  For our study we take $f_i = .01$  for every signature except
the total number of SUSY events, for which we take $f = .15$.

In order to figure out the value of $(\Delta S_{AB})^2$ that defines the typical size of statistical fluctuations, we ran a subset of our models again with a different
random number seed and calculated the $\Delta S^2$ values between duplicated models.    We define the cutoff as the 95th percentile of
these $\Delta S^2$ values, yielding $(\Delta S_{\rm 95th})^2 = 0.285$.

One concern in using a $\chi^2$-like variable to distinguishing models
is that it does not account for the fact that if one signature differs
by a large ($> 5 \sigma$) amount, the overall $\chi^2$ can be still be
very small.  This is especially a concern in our case where the total
number of signatures---1808---is very large, and we discuss this concern in more detail in section \ref{sec:really}.  One could try using the
condition that models are considered the same only if every signature
is within $5 \sigma$, however, using this criteria would force us to
reject identical models generated with different random number seeds
where one or two signatures differ by more than $5\sigma$ because of a
large statistical fluctuation.   Another deficiency of a $\chi^2$-like variable is that it does not
account for the fact that certain signatures are better than others at
distinguishing between models.  However, in trying to develop an
optimally weighted $(\Delta S_{AB})^2$ variable, we have found no
effective weighting strategy which will enhance the statistical
significance of the difference between degenerate models.

In order to quantify the distance between two models in parameter
space, we define $\Delta P^2$ as:
\be
(\Delta P_{AB})^2 = \frac{1}{n_{\rm para}}  \sum_i \left(\frac{p^A_i -
p^B_i}{\bar{p}^{AB}_i}   \right)^2, \qquad \bar{p}^{AB}_i =
\frac{p^A_i + p^B_i}{2}
\ee
where $p^A_i$ ($p^B_i$) is the value of the $i$-th parameter for model
$A$ ($B$) and the sum
runs over $n_{\rm para}$ parameters.  Roughly speaking, $\Delta P^2$
gives the quadrature
average of the percentage difference between model parameters.
Depending on the context, this sum can run over all the parameters or
just a subset.

\section{Characterizing the Degeneracies}
\label{sec:character}

The pigeonhole principle argument tells us that there must be
degeneracies, but it does not tell us what the degeneracies actually
are. And it is clearly challenging to find an algorithm to
systematically find all degeneracies associated with a given model,
precisely because they are not ``close" to each other and there is no obvious
way to continuously travel between degenerate pairs.  Plus, the existence of cliffs shows that it would be difficult to find local minima in signature distance because of the strong sensitivity of LHC observables to the SUSY parameters.  Nonetheless, as
we will see the degeneracies have rather simple characterizations,
and understanding these will help in devising strategies for breaking
the degeneracies with more signatures.

\begin{figure}
\begin{center}
\includegraphics[scale=0.9]{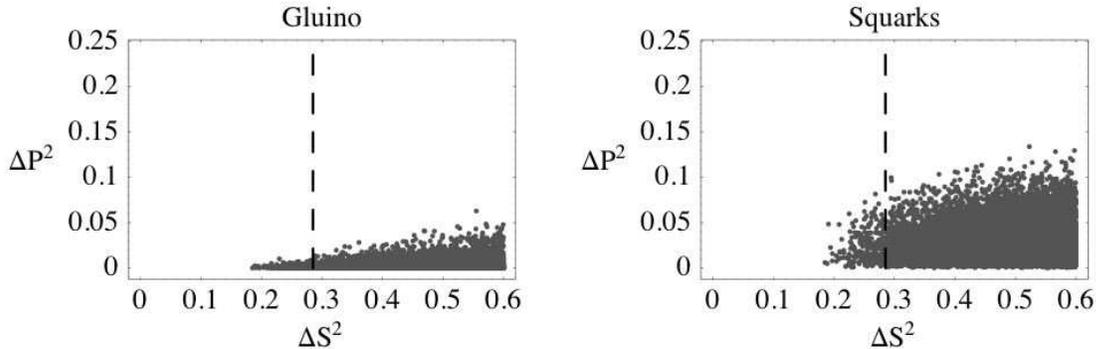}
\end{center}
\caption{A $(\Delta S^2, \Delta P^2)$ plot for gluinos and squarks,
  where again $\Delta P^2 \sim 0.01$ roughly corresponds to 10\%
  accuracy in the desired parameters.   We see that the maximum
  variation in gluino and squark masses is roughly 10\% and 40\%,
  respectively.  While not shown in these plots, the variation in the
  mass of the lightest squark (regardless of identity) never exceeds
  $\sim 15$\%. In the plot for squarks, the variation of the mass
  parameters is calculated as the average variation of individual
  squark mass parameters. } 
\label{fig:svsp-colorx}
\end{figure}

The first important point is that the gross properties of the
colored superpartner spectrum---the masses of the gluinos and
squarks---are largely determined by our signals. In figure \ref{fig:svsp-colorx} we
examine the $(\Delta S^2, \Delta P^2)$ plot where $\Delta P^2$ only includes the gluino mass separation.   This perfectly resembles the ``best of all possible worlds" plot from figure \ref{fig:vs-best}---it is beautifully triangular and there are no degeneracies for gluino
masses. The analogous plots for squarks are also shown in figure \ref{fig:svsp-colorx}.
Note that there is greater variation in $\Delta P^2$ for small $\Delta S^2$, indicating that LHC observables are not as sensitive to individual squark masses as gluino masses.  One of the reasons why the variations in squark masses are small is that squarks are scanned in a small range compared to sleptons and electroweak-inos.  However, the mass of the lightest squark is still well constrained, suggesting that the jet signatures fix some overall scale for the squarks but not their flavor or handedness.

\begin{figure}
\begin{center}
\includegraphics[scale=0.9]{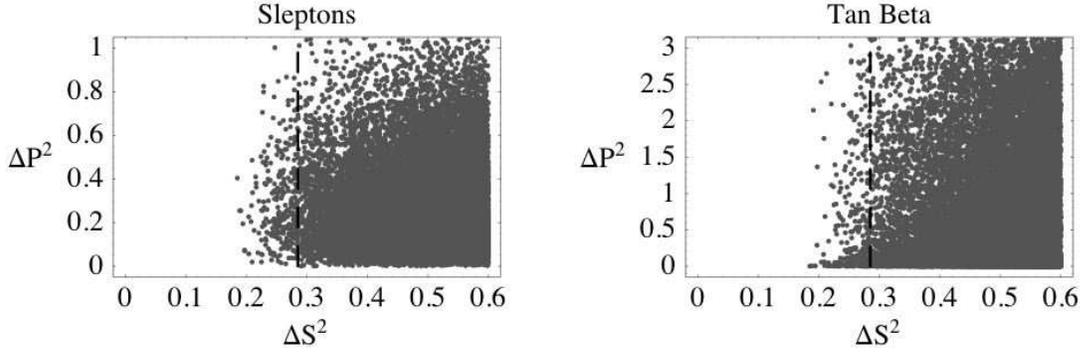}
\end{center}
\caption{A $(\Delta S^2, \Delta P^2)$ plot for sleptons and $\tan
  \beta$.  We see that none of these parameters are particularly well
  constrained. The variation of the slepton mass parameters is the
  average variation of individual slepton masses. }
\label{fig:svsp-slepton}
\end{figure}

The plot for slepton masses in figure \ref{fig:svsp-slepton} is very different.  Note that there is a continuous spread of slepton masses at small
$\Delta S^2$.  Evidently we are not particularly sensitive to the
slepton mass over large ranges of parameter space; this is
unsurprising because sleptons are not copiously produced in cascade decays
in most of the regions of parameter space we have simulated.
Similarly, $\tan \beta$ is generically unconstrained, because while
the higgsino couplings are controlled by $\tan \beta$, movement of the
third generation squarks can often compensate for changes in the
branching ratio to higgsinos.   

\begin{figure}
\begin{center}
\includegraphics[scale=0.9]{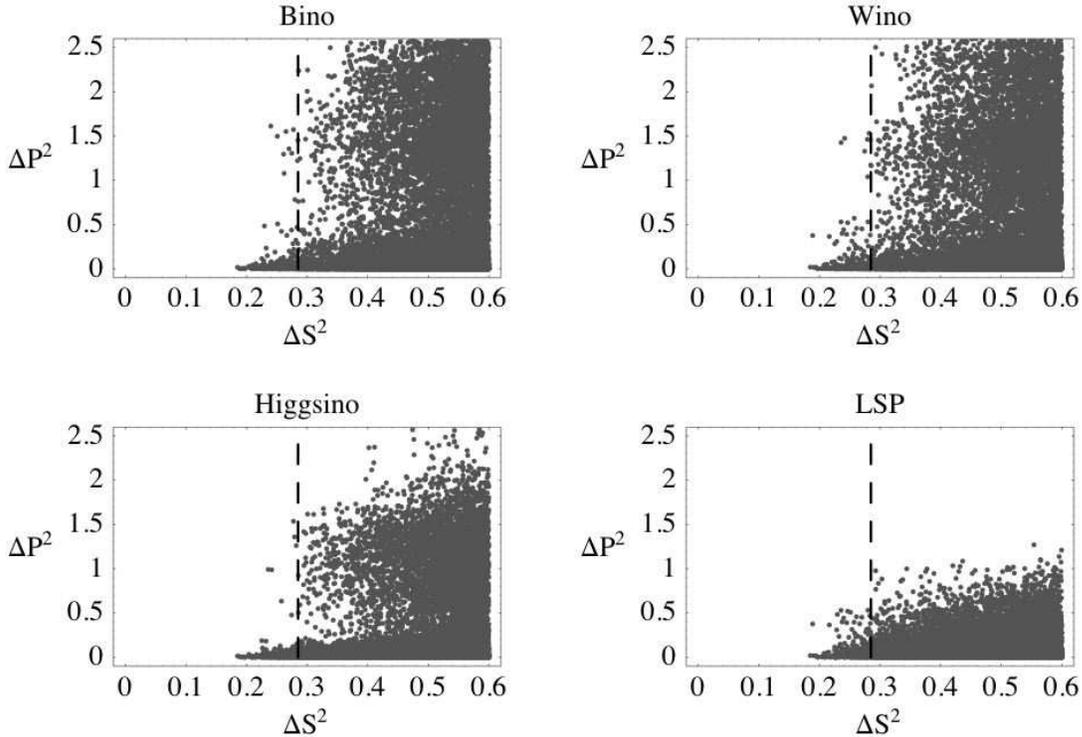}
\end{center}
\caption{A $(\Delta S^2, \Delta P^2)$ plot for electroweak-inos.  In
  these plots we see the presence of two regions at small $\Delta
  S^2$.  There is a lower triangular region, in which the
  electroweak-inos are constrained to at most 20\% - 40\% variation.
  Then there is an upper region where the electroweak-ino masses
  appear unconstrained.  This indicates the presence of discrete
  choices in the electroweak-ino spectrum.  Note that the mass of the
  LSP mass is much better constrained than the mass of the individual
  electroweak-inos, showing that LHC signatures are sensitive to mass
  eigenvalues but not mass eigenvectors.} 
\label{fig:svsp-electroweakx}
\end{figure}

The situation is different still with the electroweak-inos, as can
be seen from figure \ref{fig:svsp-electroweakx}. Here we see dramatic evidence for the existence
of degeneracies---at small $\Delta S^2$, the plot has two branches with
small and large $\Delta P^2$. This is strong evidence for discrete
ambiguities in the determination of electroweak-ino parameters.

The existence of degeneracies in the electroweak-ino sector may seem
counter-intuitive. After all, the electroweak-ino sector in the MSSM
is made of particles from different
representations of $SU(2)_L \times U(1)_Y$. Therefore, they couple
very differently to matter multiplets and to standard model gauge
bosons. For example, the bino will couple to both left-handed and
right-handed states, while the wino only has left-handed couplings.
The wino and the higgsino have both charged and neutral states---approximately
degenerate in mass---as part of
the same multiplet, while the bino only has a neutral state.

Therefore, one might expect that LHC signature should
be very sensitive to not only the masses, but also the identities of
the electroweak-inos. Indeed, two otherwise identical models with
different electroweak-ino mass parameters produce quite different
LHC signatures.  However, our key observation is that such changes of the
electroweak-ino mass parameters can sometimes be compensated by
changes of other soft parameters.

In general, we can attribute the existence of degeneracies to the
following facts.   First, in principle, the identity of final state quarks
from SUSY decay chains carries a lot of the
information about various intermediate superpartner states. On the
other hand, except for the partial flavor tag of the third generation,
all other information of the quantum numbers of final state quarks are
lost. Second, most of the kinematical observables, 
such as $P_T$, are only sensitive to the mass splitting of the
superpartners, not their identities. Third, there are large region
of parameter space where the electroweak-inos have nearly identical decay
modes, hence very similar signatures. For example, as long as the
mass splitting between the electroweak-inos is greater than $m_W$ and
sleptons are decoupled, the dominant decay mode of a chargino is 
almost always $\chi^{\pm} \rightarrow W^{\pm} + \mbox{LSP}$,
relatively insensitive to the mixings of two chargino states.
Finally, we do expect to get a better handle on the identity of the
electroweak-inos if the decays through on-shell sleptons have
significant branching ratios, due to the fact that leptonic
signatures typically carry much more information comparing to
jet signatures. On the other hand, without any theoretical
preference, this is a very special corner of
the MSSM parameter space.

Notice that overall event rate, while the most statistically
significant observable for
measuring mass scales of colored particles, is not a sensitive
observable to the more subtle structure of degeneracies involving
electroweak-inos and sleptons. Nor is it sensitive to individual
movements of squark mass parameters, as long as some rough overall squark mass
scale is fixed.

The electroweak-ino degeneracies can be characterized in a number of
simple ways, which we describe in the following subsections. 

\begin{figure}
\begin{center}
\includegraphics[scale=0.80]{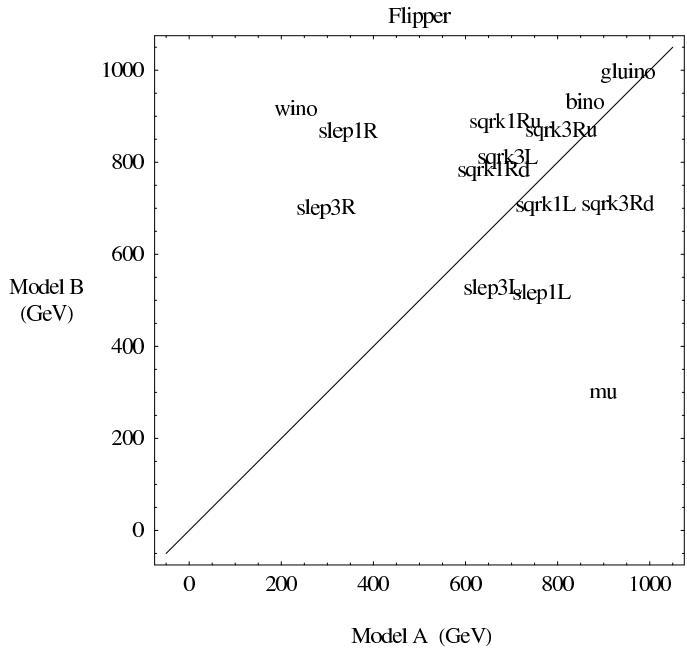}
\end{center}
\caption{An example of a Flipper, where the masses of the electroweak-inos stay roughly fixed but their identity changes.  In this example a wino LSP is replaced by a higgsino LSP. \label{flipper1}}
\end{figure}

\begin{figure}
\begin{center}
\includegraphics[scale=0.80]{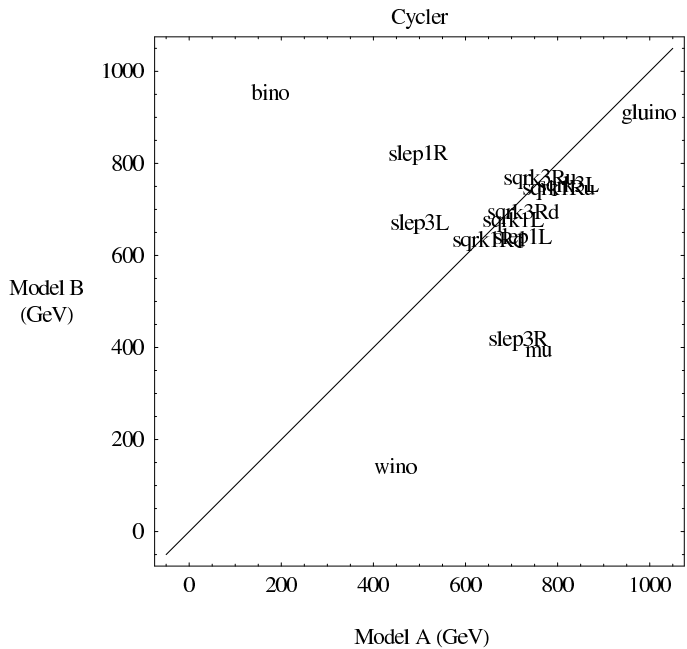}
\end{center}
\caption{An extreme example of a flipper is a Cycler, where the electrweak-ino identities undergo a cyclic permutation.  In model $A$, $\tilde{B} < \tilde{W} < \mu$, whereas in model $B$, $\tilde{W}< \mu<\tilde{B}$.  \label{cycler}}
\end{figure}

\subsection{Flippers}
\label{sec:flippers}


Flippers are probably the most dramatic example of degeneracies in
the electroweak-ino sector. In this case, although the mass
eigenvalues of two sets of electroweak-inos remain approximately the
same, the identities of two members of them are swapped. We observe that it is possible to make such a swap without
significantly changing the signatures by adjusting somewhat the
other SUSY mass parameters at the same time, such as left- and right-handed sfermions
masses.

A simple example of a flipper is shown in figure~\ref{flipper1}. In this
pair of models, although the LSPs have approximately the same mass,
their identities are wino and higgsino, respectively. Besides the LSP,
other electroweak-inos are at most barely in the decay chain,
because of  the suppressions either from phase space or from an off-shell
squark.  The dominant channel of decay of gluino and squarks are
directly to jets and LSP. Since we do not measure the charge of the
jet, such signals reveal very little about the identity of the
LSP. Notice also that the masses of the other squarks and sleptons moved in this
example, compensating possible differences from a simple swap of the wino and higgsino.


Another more dramatic example of a flipper---a ``cycler''---is shown in
figure \ref{cycler}. In this case, the identities of the three
electroweak-inos  in the degenerate models differ by a cyclic
permutation. In each of these two models, there are two
electroweak-ino states lighter 
than the gluino and squarks, and hence both are present in the decay
chain. On the other hand, in the absence of a significant slepton
branching ratio, the decay between the two electroweak-ino states is
dominated by $W$s, $Z$s and 
Higgses, and Higgs information is always less significant due to
backgrounds, trigger bias, and tagging efficiency.  Therefore, again,
the lower stage of 
decay chain, which is the most relevant for understanding the
electroweak-inos, are not very telling in this case.  We will discuss this example in more detail in section \ref{sec:really}.

\begin{figure}[p]
\begin{center}
\includegraphics[scale=0.80]{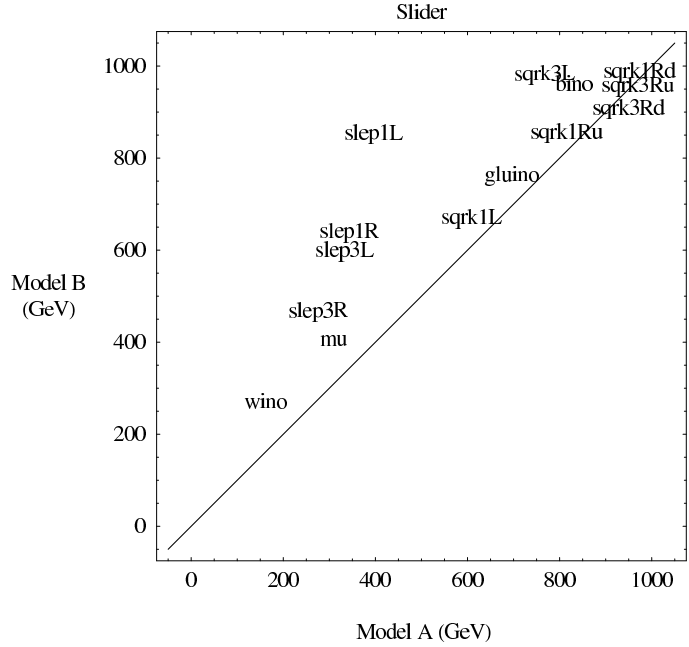}
\end{center}
\caption{An example of a Slider, where the spectrum is shifted up,
  keeping the mass difference among the electroweak-inos and gluinos
  roughly fixed.  In this case, the shift is about $100 \GeV$ for the
  most relevant parameters. \label{slider}} 
\end{figure}

\begin{figure}[p]
\begin{center}
\includegraphics[scale=0.80]{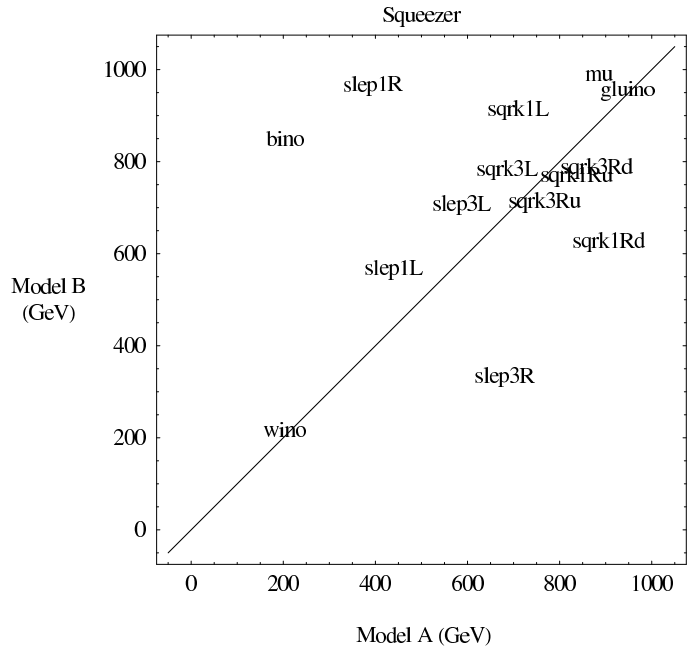}
\end{center}
\caption{An example of a Squeezer, where the mass difference between
  electroweakinos is made small such that the decay products between
  adjacent states are too soft to detect.  In this example, the LSP
  mass is fixed, but the number of neutralino states at low energies
  changes from 2 in model $A$ to 1 in model $B$. \label{squeezer}} 
\end{figure}

\subsection{Sliders}

Due to the existence of a neutral massive particle escaping the
detector, most of the LHC observables for SUSY are only sensitive to
mass differences. Therefore, we would expect that the signatures would not change significantly if we shifted the mass spectrum while holding mass
differences fixed.  Because we also use the total rate as one of the observables, we
expect that we will not be able to change the parameters a lot this
way. The slide in the spectrum will have to be somewhat uneven in different
parameters as well.

An example of such a slider is shown in figure~\ref{slider}. The
dominant production channel for these two models are gluinos. We see that
the masses of a number of states change collectively in one direction,
keeping the most important mass gaps, such as between gluino and the
LSP,  approximately fixed.

\subsection{Squeezers}

If the mass separation between the two states in the decay chain is
very close, the decay processes between them will only generate very
soft objects which are below detection threshold. In this case,
we expect to have little ability of telling them apart as different
states.

One example of such a squeezer is shown in figure~\ref{squeezer}. In
model $A$, a wino state is almost degenerate with a bino state, while in
model $B$, only a wino state light with about the same mass.  One would
expect to be able to tell the difference between these two models
since the lower lying states have different ratios of neutral and
charge states. On the other hand, it turns out, due to the structure
of other soft parameters in these two models, the dominant decay to
the light states only contain jets. Therefore, because of the charge
blindness of the jets, we lose information about the charge of the
final states.

\section{``Are There Really Degeneracies?''}
\label{sec:really}

Strictly speaking, we base our analysis and counting of degeneracies
 on a global $\chi^2$-like variable. 
 It is well-known that doing a $\chi^2$ fit with a large number of
 observables---1808 in 
our case---will be sometimes misleading. In particular, it is in
principle possible that two models differ by a large amount in a few
observables, hence are distinguishable, but can still have a small
 $\Delta S^2$ difference and be mistakenly 
identified as a degenerate pair.  This is certainly a valid
 concern, but we argue that this effect is not very important in our
 counting and characterization of degeneracies.  

First of all, the limitations of a $\chi^2$ fit are only apparent when
there are a small number of observables  that are dramatically
different.  The sort of situation that can arise is that two models
agree on everything except one signature which is different by, say,
$10\sigma$.  However, this is not likely to happen in our study---from the
analysis of section 
\ref{sec:dimsig}, we find that most of our observables are correlated
and the actual number of independent signature is small. More
specifically, if we assume there are  
$\sim 10$ independent observables in the set used in this study, then
if we compare two models and one signature is off by $5\sigma$, then
we would generically expect $10 \%$ of all the observables to also
show significant deviations, making it all but impossible for that
pair of models to have a small enough $\Delta S^2$ to be called
degenerate. 

We can also examine the degenerate examples we have found and
check for such an effect.  We can compare, for instance,
the maximum deviation in the signatures with dileptons, with the same
deviation from simulating the identical model with different random
number seed.  Dilepton signatures are chosen here since they have very
little standard model background after cuts and a significant
difference in them will definitely break the degeneracy. 

In particular, we ask what fraction of the degenerate candidates
in our database have at least one  dilepton signature deviation
greater than 5$\sigma$, and whether that fraction is consistent with
statistical fluctuations.  Out of the 283 pairs in the general scan within the strict criteria of $\Delta S^2 < 0.285$,
46 had some dilepton signal that differed by more than 5$\sigma$,
yielding a failure rate of 16\%.  But out of 2600 models run with
different random number seeds, 611 had some dilepton signal greater
than $5\sigma$, yielding a statistical expectation for a 23\% failure
rate.  Therefore, we see that the failure rate is consistent with
statistics, showing that generically our definition of $\Delta S^2$
identifies true degeneracies. 


We could do such a comparison in other classes of signatures as
well, but such a comparison will be less meaningful for
several reasons.  Signatures with fewer leptons typically
have much larger standard model background. Therefore, the statistical
errors we 
have used in our study are not realistic.  In particular, we expect only large and qualitative differences in these signatures to be useful in distinguishing SUSY models. At the same time, compared with dilepton
signatures, it is much easier to make small
changes in the soft parameters to obtain a better fit to zero or one lepton
signatures. For example, a large deviation in the total number of SUSY
events can be compensated by a very small change in the gluino mass with
negligible effects on the other signatures. 

To give a sense of how different the LHC observables are in a typical
degenerate pair of models, we can look more closely at a specific
example which has interesting 
swaps in electroweak-ino masses---the ``cycler'' example shown in section
\ref{sec:flippers}.  There is no significant
difference in the dilepton signatures for the cycler pair. Therefore, it is not at all
obvious how we should distinguish them. This is also a pair of models where we
do see some discrepancies in some of the other signatures, which are however unlikely to be useful to
qualitatively distinguish the two models.  

\begin{figure}
\begin{center}
\includegraphics[scale=0.9]{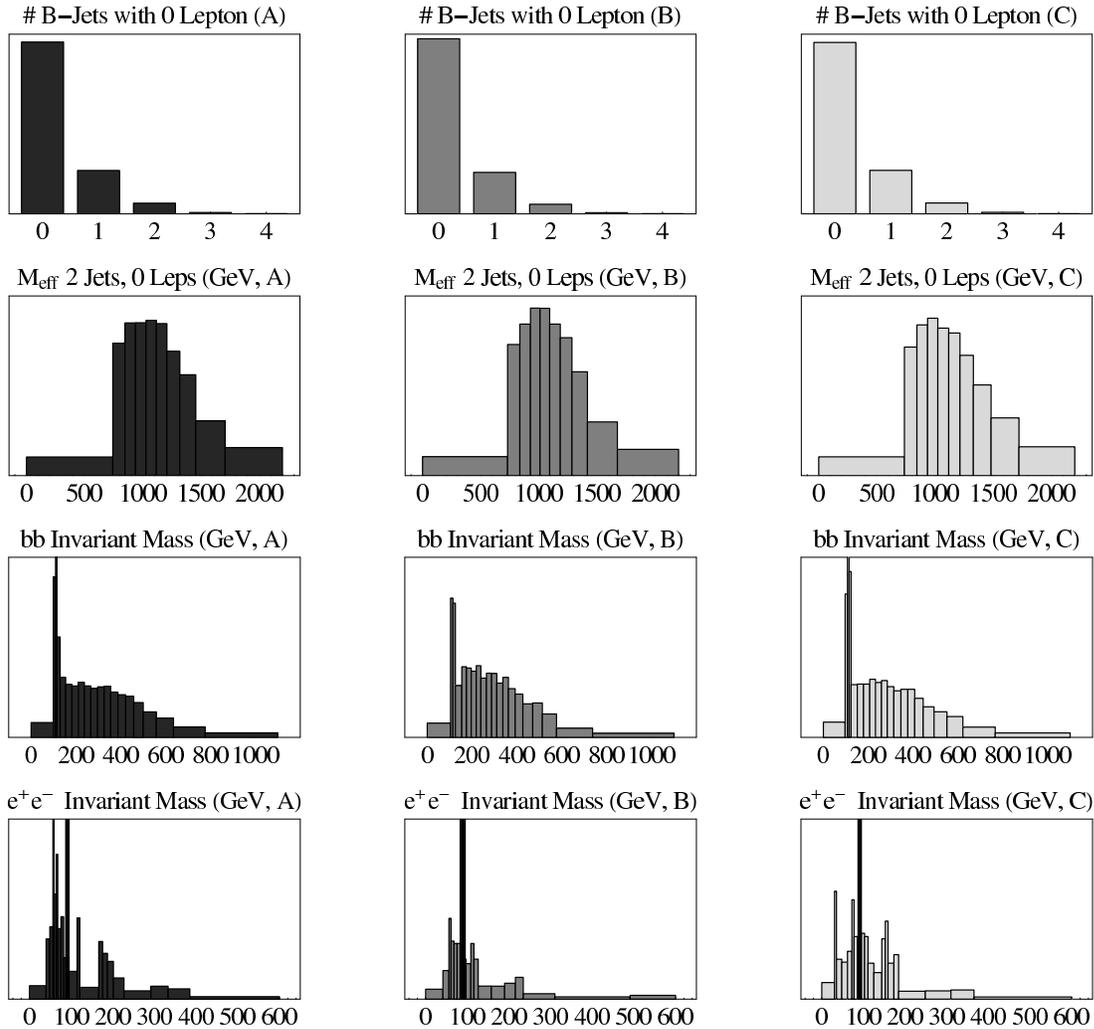}
\end{center}
\caption{Example distributions from the ``cycler'' example of figure
  \ref{cycler}.  Model A and B are a degenerate pair, while model C
  has the same parameters as model A but was generated with a
  different random number seed.  Note that histograms are plotted in
  quantiles---equally occupied bins of variable width.   With the
  exception of lepton charge and total number of events as discussed
  in the text, all counting signatures have roughly the same variation
  among the three models as the $b$-jet counting example shown here.
  Similarly, all kinematic histograms have the same or smaller
  variation as the $M_{\rm eff}$ example.  While the dielectron invariant
  mass distribution looks different between models $A$ and $B$, it
  also looks different between models $A$ and $C$, so we can ascribe
  such differences to statistical fluctuations.  The main difference
  between models $A$ and $B$ is seen in the rate of Higgs production
  from the $bb$ invariant mass distribution, though as we argue in the
  text, such a difference is washed out by $t\bar{t}$ standard model
  background and also depends on the details of the Higgs sector,
  which we made no effort to control in our study. 
} 
\label{cycler-dist}
\end{figure}

A set of typical observables for this pair are plotted in
figure~\ref{cycler-dist}.  Columns labeled by A and B are from the
degenerate pair of models. Column C shows distributions from running
model A again with a different random number seed, which provides a
guide for the size of statistical fluctuations. Visually, we see that there
is no difference in the counting of $b$-jets or in the $M_{\rm eff}$
distribution, despite the fact that those distributions showed some
variations between models A and B ($1\sigma$--$3 \sigma$).   Except
for the signatures we discuss below, all other counting signatures and
distributions showed comparable or smaller deviations.    

The shapes of the dielectron invariant mass distributions are somewhat
different. On the other hand, the difference between models A and
model B are at 
least as big as the difference between model A and model C, showing
that at 10 fb$^{-1}$, there is not much information in these dilepton
invariant mass distributions.  Whether there could be some smoking gun
in this distribution at 300 fb$^{-1}$ is unknown. 

One potential difference in these two models is the $b \bar{b}$
invariant mass distribution. In both models, Higgs bosons are
produced as part of the decay chain. In model A, Higgses are
more copiously produced, and there is a 4$\sigma$ difference in the rate
of Higgs production. On the other hand, it is very unlikely such a
signature will be useful, even at high luminosity. First of all, at
least with our cuts, 
such a deviation will not be visible once we include the standard
model $t \bar{t}$ background. Moreover, such a difference is highly
sensitive to the details of Higgs sector. If there is a significant
modification of Higgs sector of the MSSM, such a including a singlet
in the possible decay product of the Higgs, this deviation will almost
certainly be most less prominent. 

The largest deviation in the observables between these two models
is a somewhat large charge asymmetry in the single lepton
signature, about $20 \%$, corresponding to $7\sigma$--$8\sigma$ using
only statistical error bars. However, distinguishing these two models based on this difference in single
lepton signature will be very challenging due to the existence of
large standard model background.  At higher luminosity, one might
expect the charge asymmetry to appear in trilepton events, but given
the absence of sleptons in the decay chain, trilepton signatures will
be relatively sparse. 

Furthermore, the charge asymmetry is not necessarily a robust
qualitative difference between these two scenarios either. The
dominant contribution to such a difference comes from squark-gluino 
associated production, which is in turn very sensitive to small
differences in squark masses.   Because the
overall rate of SUSY events between models A and B differ by $4
\sigma$ due to a 50 GeV shift in the gluino mass, we could fine-tune
the gluino mass in one of the models to make the overall SUSY
production rate more comparable, and such a fine-tuning (accompanied
by other shifts in the squark sector) could alleviate the charge
asymmetry without drastically modifying other observables.

The cycler example is chosen here not because the pair has completely
identical signatures, but because it exemplifies an
interesting structure of degeneracy. The fact that two such
drastically different models could come this close with observables
extracted from only pure signal gives us the confidence that they are
good representatives of a degenerate scenario. 

Because of the ambiguity of the $\chi^2$ analysis, there is no reason
to believe that the best examples of degeneracies should be within the
signature cut of $\Delta S^2 = 0.285$. There are
pairs with somewhat larger signature separation and yet with no
qualitative difference in any particular signature. The apparent
difference in total $\chi^2$ comes from adding up small deviations
from a number of observables, which is expected due to the large
number of observables. Such 
pairs may actually be a better representative of a qualitative
degeneracy, and we expect that nearby models in parameter space would form degenerate pairs satisfying a stricter degeneracy criteria.

\begin{figure}
\begin{center}
\includegraphics[scale=0.80]{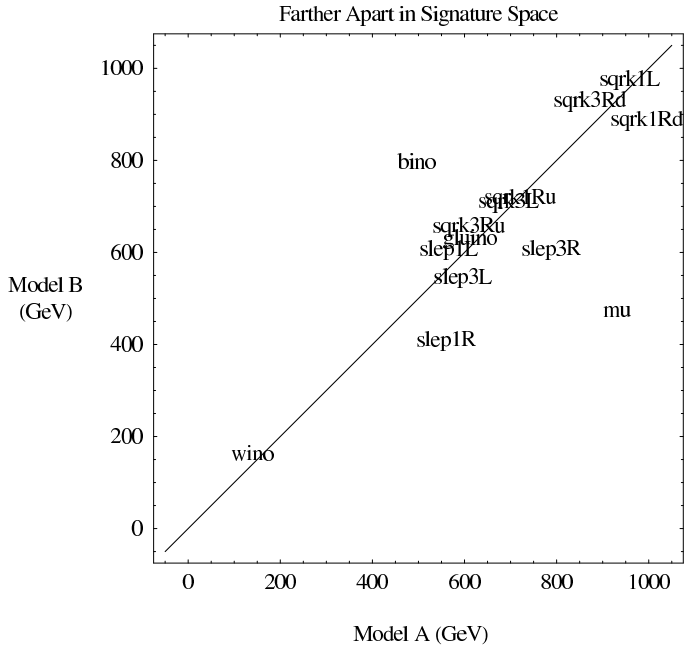}
\end{center}
\caption{An example of a pair of models whose signature distance
  ($\Delta S^2 = .457$) is larger than the cutoff defined in this
  paper for degenerate models.  While these models can be
  distinguished by our $\chi^2$ fit, it is possible that by slightly
  changing the parameters in this model, a robust degeneracy could be
  formed.  This pair qualifies as a flipper, in that the identity of
  the second electroweak-ino is changed from a bino in model $A$ to a
  higgino in model $B$.\label{fig:pointfive-mass}} 
\end{figure}

\begin{figure}
\begin{center}
\includegraphics[scale=0.9]{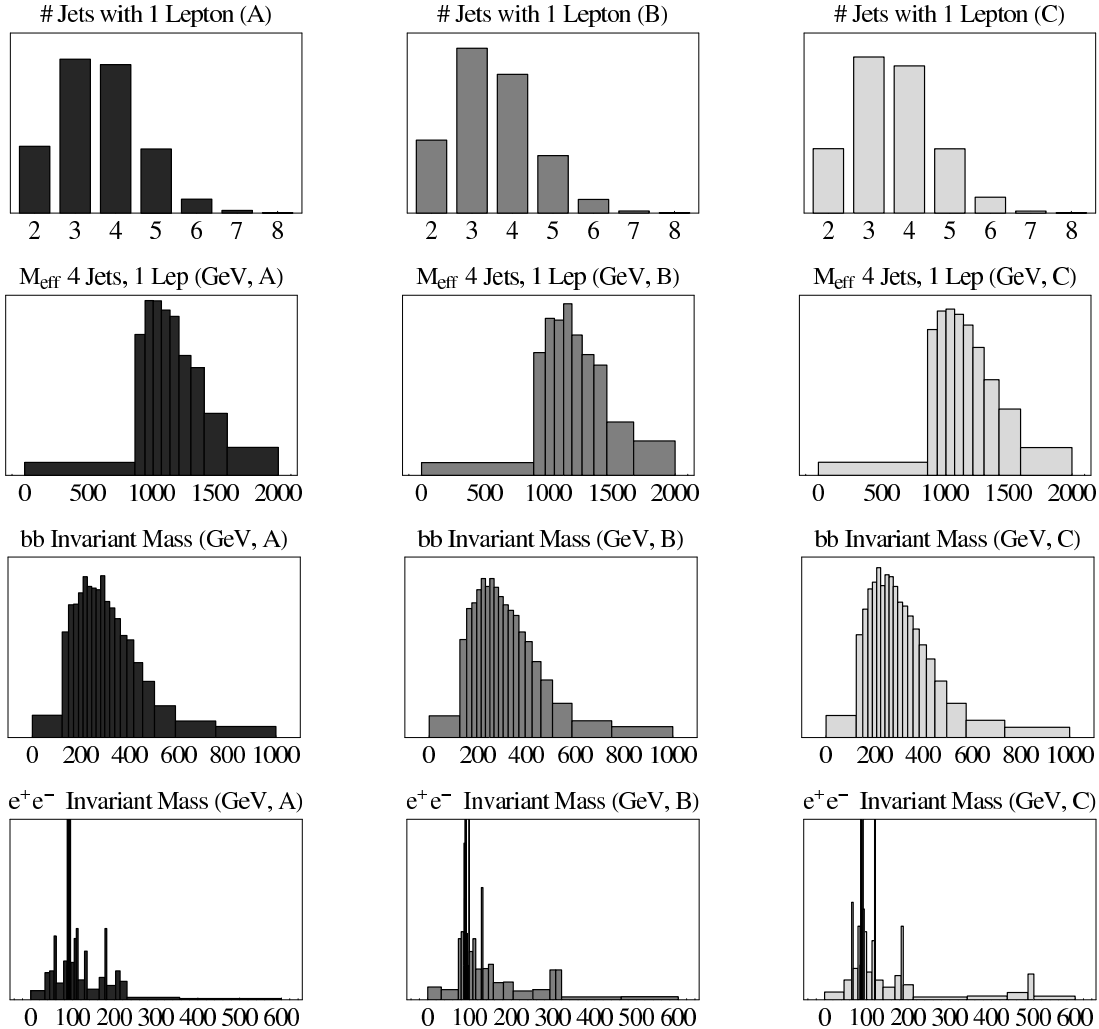}
\end{center}
\caption{Example distributions from the pair of models in figure
  \ref{fig:pointfive-mass}.  Like figure \ref{cycler-dist}, model $C$
  has the same parameters as model $A$ but was generated with a
  different random number seed.  The most noticeable difference in
  these plots is a slight imbalance in the number of 3 jet vs.\ 4 jet
  events with 1 lepton present.  All of the other distributions are
  qualitatively similar between models $A$, $B$ and $C$; a typical variation is shown for a $M_{\rm
    eff}$ plot.  While a slight bump in the dielectron invariant mass
  distribution is present in model $B$ at 300 GeV, the significance of
  such a feature is called into question by the presence of a similar
  bump in model $C$ at 500 GeV which is entirely absent from model
  $A$, despite the fact that they share the same underlying parameters.}  
\label{fig:pointfive-plots}
\end{figure}

Such an example is shown in figures \ref{fig:pointfive-mass} and
\ref{fig:pointfive-plots}.  This pair of models is a flipper
degeneracy where the second-lightest electoweak-ino changes from a
bino to a higgino.  While there are many $1\sigma$ variations in
numerous signatures, there is no qualitative difference between these
two models in any of the inclusive signatures. The most noticeable
difference is an $8 \%$ difference in the inclusive one lepton
counting, whose effect is somewhat visible on the plot in figure
\ref{fig:pointfive-plots}.  There is also a $9 \%$ 
difference in the counting of 2 jet with no lepton events. Such
deviations will almost certainly be swamped by large standard model
background.  There is no charge 
asymmetry in one lepton events, nor evidence for different rates of
Higgs or $Z$ production, nor any other potentially qualitatively
different structure in all the other signatures.  Such an example
suggests that in the vicinity of pairs of models that lie a moderate
signature distance apart, one could arrange robust degenerate pairs by
making small parameter variations to smooth out $1\sigma$
differences. 

\section{Sleptons and Long Cascade Decay Chains}
\label{sec:sleptons}

Part of some people's optimism about making precision measurements at the LHC
comes from studying leptonic signatures.  Not only is the standard
model background smaller for events with hard jets and missing energy
and hard leptons, but leptons, with their charge and flavor
identified, carry much substantial information about the underlying
processes. Moreover, the energy resolution on electrons and muons is
much better than for jets.  So while colored particles in SUSY
guarantee that SUSY cross sections will be large at the LHC, it is
leptons from cascade decays that are entrusted with constraining the
SUSY spectrum.

In constrained models like mSUGRA or GMSB, we have reason to believe
that slepton will be light, and therefore a reason to expect
electroweak-ino-slepton-electroweak-ino cascade decays. Notice that
the parameter region where we have a large slepton production in the
decay chain is quite special. It certainly requires electroweak-inos,
in particular those with large gaugino fractions, to be heavier than
the first two generations of sleptons. At the same time, it also
requires those inos to be  light enough so that they are in the decay
chain from gluinos and squarks.  This set of requirements enforce
a well-ordered mass spectrum from a general MSSM point of view. 

\begin{figure}
\begin{center}
\includegraphics[scale=0.7]{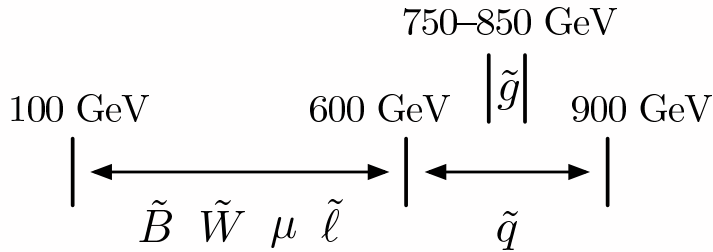}
\end{center}
\caption{The parameter ranges for our dedicated cascade decay scan.
  Unlike the previous scan, we assume flavor universality for the
  sleptons and squarks, but we will still split left- and right-handed
  sfermions, and up- and down-type squarks.   We force that the heaviest slepton is lighter than at
  least one electroweak-ino.  As before, $\tan \beta$ ranges from 2 to
  50.} 
\label{fig:mass-cascade}
\end{figure}

In order to explore this special situation better, we have conducted a
dedicated scan, as shown in figure~\ref{fig:mass-cascade}, in
which we enforce the existence of some long decay chains.
In particular, the heaviest slepton is required to be lighter than at
least one of 
the electroweak-inos.  Also, we do not split the third generation of
squarks and sleptons from the first two generations, with the idea
that in most theoretical models with long cascade decays, there is
usually some kind of flavor universality. 


We simulated $m = 27196$ models, and still found a number of
identical doubles, $N_2 = 56$. This gives us an estimate for the
total number of bins in signature space
\begin{equation}
N_{sig} \sim m^2/N_2 \sim 6.6 \times 10^6
\end{equation}
which is somewhat larger than in our previous example, despite the
smaller range of parameter space scanned. This clearly indicates
that new leptonic directions in signature space are being opened
up and we should expect fewer degeneracies.

Indeed, of the 56 degeneracies we found, {\it none} correspond to
reasonably ``different" models, which by equation (\ref{eq:degen}) yields
\be
\langle d \rangle \sim 1.
\ee
The pigeonhole principle argument
also gives us an estimate of the number of degeneracies in the
cascade decay with on-shell sleptons.  Suppose we could have the
following accuracies on the measurement of soft parameters: gluinos to
50 GeV, both left- and right-handed squarks to 75 GeV, both left- and
right-handed sleptons to 75 GeV, and electroweak-inos to 50 GeV. This
gives us\footnote{Again, these values are chosen by estimating the local variation in mass parameters from $(\Delta S^2, \Delta P^2)$ plots for our cascade run.  However, the separate plots for gluinos, squarks, sleptons, and electroweak-inos are not shown in this paper.} 
\be
N_{\rm models} \sim 2 \times 4^3 \times 6^2 \times 10^3 \sim 5 \times
10^6 \sim N_{\rm sig}, 
\ee
So by equation (\ref{eq:degenalt}), we should be left
with some ${\cal O}(1)$ number degeneracies in this case.

\begin{figure}
\begin{center}
\includegraphics[scale=0.9]{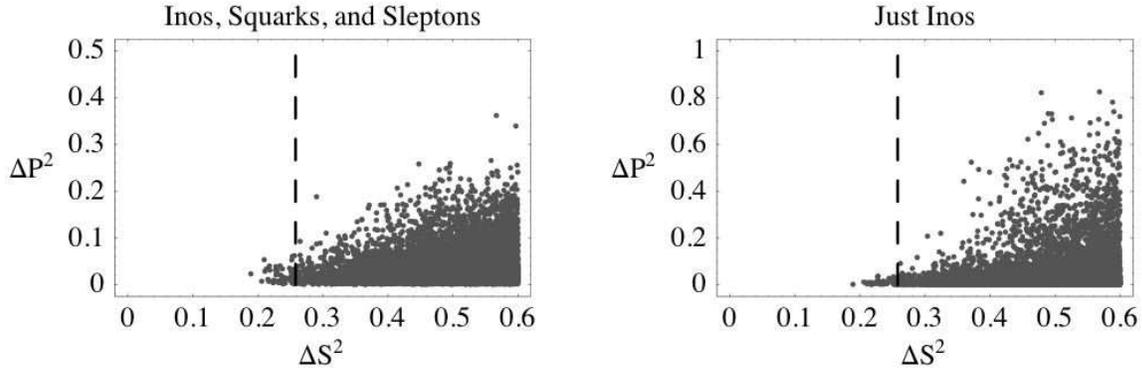}
\end{center}
\caption{A $(\Delta S^2, \Delta P^2)$ plot analogous to figure
  \ref{fig:svsp-master}, where we see no evidence for degeneracies.
  The $\Delta S^2$ between identical models is now $\Delta S^2 =
  0.258$.  The left plot indicates that mass parameters can shift by
  at most 20\%.  The right plot shows no evidence for bimodal behavior
  in the electroweak-ino sector.} 
 \label{dp_ds_sl}
\end{figure}

\begin{figure}
\begin{center}
\includegraphics[scale=0.9]{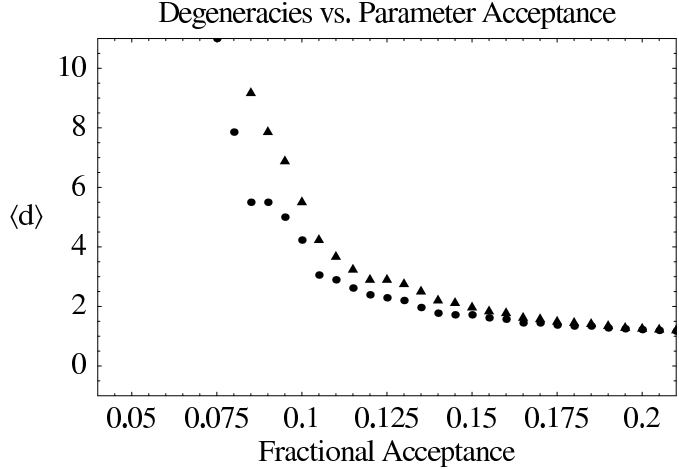}
\end{center}
\caption{Number of degeneracies as a function of allowed fractional
error in cascades run.  Triangles correspond to imposing the fractional
acceptance on gaugino, higgsino, and squark parameters.  Dots
correspond to just imposing fractional acceptance on the inos.  We see
that the number of degeneracies asymptotes to 1, indicating that the
only ``degeneracies'' we have observed are coming from uncertainties in mass
measurements and not from discrete choices in the spectrum.} 
  \label{degen_percent_sl}
\end{figure}

The reduction of the number of the degeneracies in this scenario can also be
be seen in the $(\Delta S^2, \Delta P^2)$ plots shown in
figure~\ref{dp_ds_sl}. More quantitatively, the number of expected
degeneracies as a function of fractional differences in the parameters
is shown in figure~\ref{degen_percent_sl}. As we discuss more below,
the reduction in the number of 
degeneracies is mainly due to the fact that with long decay chains, it is in
general much harder to find flippers in the electroweak-ino
sector. Notice also that the number of degeneracies 
including squarks in figure~\ref{degen_percent_sl} is also much
smaller compared with our general scan. This is partly due to the fact
that we have chosen the 
three generations of soft masses to be universal, but equally
important, our better handle on the identities of the electroweak-inos
from sleptons translates into a better handle on the left-right splitting of the squarks. 

\begin{figure}
\begin{center}
\includegraphics[scale=0.45]{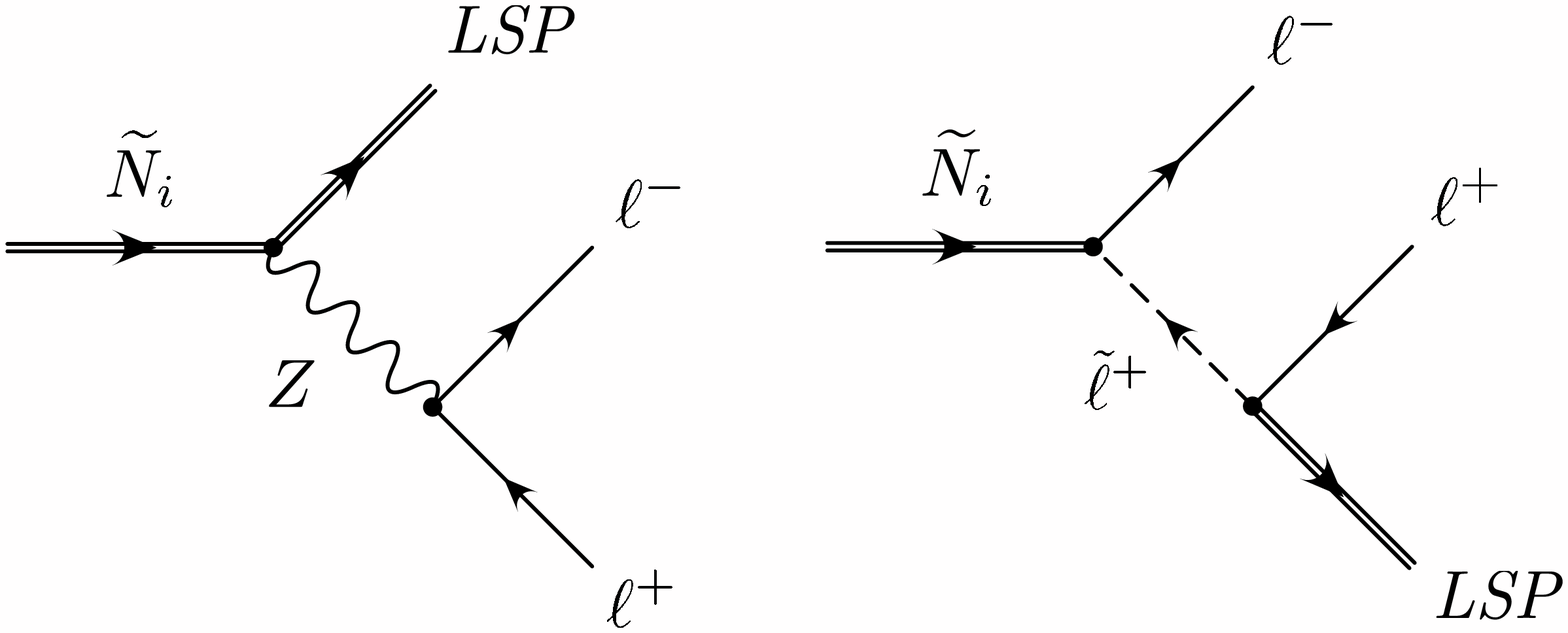}
\end{center}
\caption{The two major processes that can lead to correlated same-flavor opposite-sign dileptons in SUSY cascade decays.  When dileptons come mostly from on-shell $Z$ bosons as in the left diagram, then crucial information about the masses and identities of the neutralinos is lost.  Only when sleptons are copiously produced in  cascade decays as in the right diagram, do we gain a strong handle on the identities and mass splittings of neutralino spectrum.}
\label{decay_sl_z}
\end{figure}

The main reason that the presence of the sleptons significantly reduces
the number of degeneracies is their role in decays of the
type shown in figure~\ref{decay_sl_z}.  Without sleptons in the decay
chain, the electroweak-ino decays are typically dominated by
$W/Z$s---the left panel of figure~\ref{decay_sl_z}---and 
Higgses.  Because $h \rightarrow b \bar{b}$ typically carries
less information due to backgrounds and tagging
efficiencies, the only robust handle on electroweak-ino decays comes
from leptonically decaying $W$s and $Z$s.   However, this handle is of
limited value, because we not only lose statistics 
because of the small leptonic branching ratio of $W/Z$, but we also lose
information about the electroweak-ino identity because the leptons are not
directly coupled to them. 

On the other hand, the situation is significantly better with (on-shell)
sleptons in the decay chain. The leptonic branching ratios as
well as the charge of the leptons are sensitive to the
identities of the electroweak-ino states in decay processes of the type
shown in the right panel of figure~\ref{decay_sl_z}.  In addition,
slepton edges \cite{atlastdr,Gjelsten:2004ki,Gjelsten:2005aw,Lester:2005je} give a strong constraint on the mass difference between
electroweak-inos, further limiting the possibilities for degeneracies. 

\begin{figure}
\begin{center}
\includegraphics[scale=0.80]{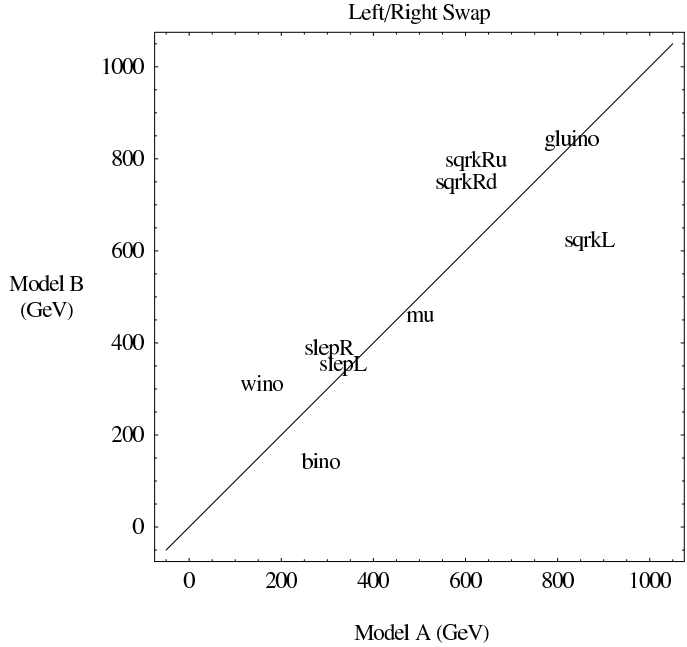}
\end{center}
\caption{An example of a Left/Right Flipper.  Though this example does
  not fall within the $(\Delta S)^2$ cut that defines typical statistical fluctuations,
  the models are sufficiently close in signature space that slight
  adjustments in the colored sector might make this a real
  degeneracy.  In this example, the left- and right-handed sfermions
  switch places to accommodate a bino-wino flip. 
 \label{swap_example}}
\end{figure}

\begin{figure}
\begin{center}
\includegraphics[scale=0.7]{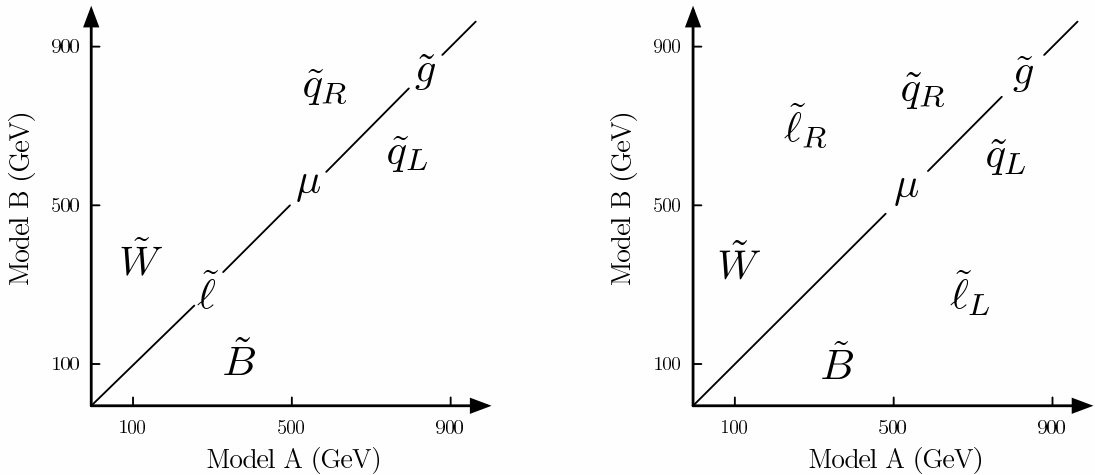}
\end{center}
\caption{Possible examples of Left/Right Flippers involving copious
  slepton production.  In order to avoid a double slepton edge from
  bino decays, either the sleptons must be nearly degenerate, or only
  one handedness of slepton can be involved in the decay.} 
\label{swap_cartoon}
\end{figure}

Still, we expect that  in some
cases it should be possible to flip the identities of the
electroweak-inos in combination with flips of the other sfermion
states. One example of such a flipper is shown in
figure~\ref{swap_example}, though strictly speaking this pair of
models does not satisfy the degeneracy criteria of section
\ref{sec:comparing}. In this 
example, the left- and right-handed squark states flip in order to
accommodate a flip 
of the bino and wino. On the other hand, this is not a perfect example,
since the sleptons are still somewhat decoupled except through small
mixings between the higgsinos and the gauginos. 

If there do exist left-right flippers involving sleptons, we expect
them to be subtle since they have to reproduce a 
large amount information contained in the leptonic signatures.  In the
case that non-degenerate left- and right-handed sleptons are produced
in cascade decays, it would be extremely 
difficult to flip the wino and bino around the slepton states; 
a scenario with heavier bino will have two edges in the dilepton
invariant mass spectrum, while a scenerio with a heavier wino will
only have one edge.  

On the other hand, there could be other scenarios in which we could
have a flipper. Two of those possibilities are shown in
figure~\ref{swap_cartoon}.  It is possible that sleptons in the decay
chain have near universal masses, as shown in the left panel. It is
also possible the left- and right- handed sleptons are playing
different roles in the bino-wino flip case, as shown in the right
panel.  In both of these cases, a double edge \cite{Hinchliffe:1998ys,Paige:1999ui}
 would not
exist, but it may be the case that these models could be
distinguished by a subtle difference between dilepton and trilepton
events. Verifying the existence of such degeneracies is a challenging task. 







\section{Future Directions}
\label{sec:future}

As mentioned in section \ref{sec:details}, we have made two main compromises in
our study of the inverse map by not including standard model
background and by only simulating $10$ fb$^{-1}$ of data. In order to fully
realistically assess the potential of measuring SUSY parameters at LHC,
these two issues must be addressed. 

Various significant components of
standard model backgrounds to SUSY signatures---such as di-boson, $t
\bar{t}$, and $W/Z+ \mbox{jets}$---have been studied for supersymmetry
searches \cite{Baer:1995nq,Baer:1995va}.  The presence of standard
model background 
will no doubt worsen the sensitivity of LHC observables, in particular
pure jet signals, to soft parameters. Therefore, it is crucial to
study the extent to which it will change the number or structure of
the degeneracies. 

Similarly, the LHC is going to achieve a much larger luminosity
than we have simulated.  With a higher luminosity of $\sim 300$~fb$^{-1}$, we could in principle 
introduce more stringent cuts to purify the signal without losing
statistical significance. Therefore, compared with our study, a full
LHC data set 
with high luminosity will significantly reduce statistical
fluctuations, and a fully realistic characterization of the
inverse map will have to take this into account as well.  In
  addition, a number of processes that can only be observed with
  larger rates might be valuable in removing degeneracies.

The list of observables in appendix \ref{sec:observelist} are by no means
exhaustive, nor are the selection cuts completely
optimized.  Our study shows that in order to break degeneracies at the
LHC it is very important to construct 
new independent observables, and it may be possible to optimize the
event selection 
criteria, or use multiple selection criteria, to increase the
effective dimensionality of signature space.  Furthermore, observables
such as $g_\mu-2$, dark matter detection and density, and perhaps $B_S
\rightarrow \mu^+ \mu^-$ depend on on the same parameters in different
ways, and might be important in removing degeneracies. They could
simply be added as additional signatures for each model.

One of the major reasons for the existence of degeneracies is the loss
of information due to the hadronization and the formation of jets. One
possible direction that deserves much more careful study is how much, 
information about the initial parton we can get out of a jet in
addition to its four-momentum, by including the information such as track
charge weighted by energy \cite{jetcharge}, or better observables for the same
physics. A measurement of jet charge with any significant 
confidence would be very effective in removing degeneracies if it were
possible. On the other hand, due to  the
large  statistics of the jet signatures, even some small preference at the
percentage level could help in principle. 

Supersymmetry events usually have a large number of jets from
different decay chains, as well as from different parts of the same decay
chains. Therefore, even with a pure signal sample, important structures
in the jet kinematical distributions are usually swamped by combinatorial
backgrounds. It will be necessary to develop  more
sophisticated methods of pairing up the correct combinations of jets
if we are to exact useful information about decay kinematics. 

Because leptonic signatures are so informative, it
would be helpful if we could get as many leptonic events as
possible. It is possible to enhance the leptonic signature by
using smarter cuts in special cases. For example, in certain scenarios,
leptons could tend to be very hard while the jets
are softer. In this case, a better selection criteria would
focus on events with hard leptons instead of imposing such strong cuts
on the jets. Careful study of using different cuts for each of two or
more leptons could be useful. 


Since we expect the sizes of degenerate islands to be small, and because the
dimensionality of the parameter space is large, it is very difficult
to scan densely and simulate a sizable sample of data at
each step. One possible method for identifying degeneracies is to
carry out a leading order 
scan by matching a limited set of important rate$\times$branching
ratios to narrow down potential regions. Such a scan will not require
event simulation and hence is much less time/resource
consuming. After such an initial scan, we could then start to scan and
simulate in a much narrower region in order to search for and identify 
degeneracies. 

A particularly interesting example of a degeneracy study would be
based on the mSUGRA model 
SPS1a \cite{sps}. This model has an on-shell slepton in the decay chain and hence
strong multi-lepton signals. Based on our study, we should have less
freedom in such a scenario to move the soft parameters around---especially the
slepton and electroweak-ino masses---without producing large
difference in observables. On the other hand, even in this case, we
still expect to see more delicate flippers which combine the flipping
of the bino and wino with shifts in the sfermion
masses. Therefore, it would be interesting to do a dedicated study
to find generic SUSY models degenerate with SPS1a. The most important
outcome of such a study 
would probably be a systematic method for mapping out degeneracies. 

Another way to understand degeneracies is to try to determine the
shapes of the small islands in parameter space.  Though we found in
our study that the size of regions in the parameter space
corresponding to a signature bin were small, if we had more details
about the shape of the small islands in each direction of 
parameter space, it would help us in mapping out the set of
solutions matching LHC data.  One such study could be based on parameter
excursions around a generic set of MSSM models and one could measure the
rate that signatures change in different parameter directions.  

Measuring the difference between left-handed and right-handed
squarks turns out to be quite difficult. We expect the overall rate will
fix roughly some overall scale of the squarks. Although the structure
in left-right squark degeneracies is much less 
prominent than the case of electroweak-inos, such a splitting is not
expected to be a flat direction either. Therefore, details of a left-right squark
degeneracy remain one of the important directions that need further
exploration. 

Of course, the statistical method we have developed in this study is
completely general and applicable to any class of models.  It is
important to repeat the same analysis not only within a class of models for new
physics but also between such classes. In particular, it will be
interesting to observe the inverse maps 
of models with similar gauge quantum numbers to SUSY, such as universal extra
dimensions or the little Higgs with $T$-parity.  Understanding robust
distinctions between those models and supersymmetry is one of the most
important questions in interpreting LHC data.  

\section{After a Discovery at the LHC}
\label{sec:discovery}

If new physics is discovered at the LHC with a pattern of signals
roughly consistent with SUSY, the most pressing challenge 
will be to invert the LHC signatures and extract information about the
soft supersymmetry breaking parameters. 

An obvious first step is finding \emph{any} model that fits the set of LHC
signatures that have been measured, presumably starting with a reduced set of SUSY parameters.  Our study
shows that the success in finding one model which fits the data is
not the end of the process, though. In generic regions of SUSY parameter
space, if there are not 
significant additions to the list of signatures we have used, there
will be different 
models which produce indistinguishable signatures.  The
number of degeneracies is not intractably large, suggesting that
degenerate models may be distinguishable with more detailed study. 

Therefore, the most important task after finding one model that fits the LHC data is to create a catalog of all models which have signatures consistent with observation.  The reason is that any detailed study would presumably require the construction of
special observables designed to distinguish specific
models.  Due to the small size of the islands in the parameter space occupied by
degenerate models, and the large volume in which those
islands are scattered, mapping out degeneracies is challenging. The
naive approach of carpeting parameter space will not work in this case
because of the large amount of necessary computation.  

Clearly, new methods and insights need to be developed for the
purpose of identifying potential degeneracies. However, our
characterization of the possible degeneracies as flippers, sliders,
and squeezers is a useful first step in finding other candidate
models.  Then again, even though there are simple characterizations of
the degeneracies in the 
electroweak-ino sector, we saw that other soft parameter had to shift
to accommodate such changes, complicating the task of manufacturing
degeneracies.  There have been some attempts to match parameters of supersymmetry breaking with LHC
data \cite{Lester:2005je,Lafaye:2004cn,Bechtle:2005qj,Allanach:2004my}, and it would be interesting to see whether these approaches can be modified for the purpose of finding degeneracies. 

\section{Discussion and Outlook}

In two years, the LHC will begin to answer questions that have driven
much of the theoretical activity in our field for the past three
decades. Thinking about how we will go from LHC data to the
underlying model {\it now} will help us get to the physics we really
care about as quickly as possible,  and may be able to improve
  experimental settings and procedure. In this paper, we have initiated
a systematic study of the LHC inverse problem, within the context of
the supersymmetric standard model with minimal particle content but
(relatively) unconstrained parameter space. We have used simple
statistical techniques to probe this map, in particular studying the
average number of models with equivalent LHC signals, and the
effective dimensionality of the signature space populated by SUSY. 

In the regions of parameter space where the sleptons are not produced in
long cascade decay chains, we find that there is very little handle
on the slepton masses, and degeneracies in the electroweak-ino
sector. The typical number of degeneracies for any given model is of
order $\langle d  \rangle \sim 10 - 100$. We have shown that these degeneracies have simple
interpretations.  For instance, in the decay of $\chi_2^0$ to
$\chi_1^0$, we can have ``flippers" where $(\chi_2^0,\chi_1^0)$ are
either $(\tilde{B},\tilde{W})$ or $(\tilde{W},\tilde{B})$, with
accompanying changes in the rest of the spectrum to match  LHC
signatures.  With sleptons produced on-shell in cascade decays as is
the case in  many of the well-studied mSUGRA models, but perhaps not
in nature, the situation is better, but 
there may still be possible degeneracies involving left/right swaps.

Our study of the inverse problem reconciles two orthogonal views one
often hears about what the LHC can determine about weak-scale
physics. There is the school of thought that says that the LHC is
only a discovery machine, but that any more precise determination of
the underlying physics must await the construction of a linear
collider. Another school of thought holds that not only can the LHC
discover new physics, it can also determine model parameters to few
percent accuracy!   Our picture of the inverse map shows the sense in
which both of these pictures can be correct. There can indeed be a
relatively large number of different models compatible with LHC
data, partially justifying the first view, but each of these is a small
island in parameter space, partially justifying the second view. Our
result for the number of degeneracies $\sim 1-100$ is as
interesting as it could have been, however.  The number is not
$10^6$; it is just large enough to represent a non-trivial challenge
and  just small enough to spur us to think of clever new signatures to
resolve the small number of ambiguities in SUSY. And it is easy to
determine whether a new set of signatures is effective in enlarging
signature space---we simply add the signature and compute the new
average number of degeneracies $\langle d \rangle$.

Obviously, the number of degeneracies will be smaller if a more
restrictive parameter space is chosen. Again, this can be looked at
by repeating our analysis for the restricted models and computing
$\langle d \rangle$. If $\langle d \rangle \sim 1$, then if such a
simple model reproduces LHC data, it is not likely to have a
degenerate pair within its own model space, and despite the fact that it
may have $\sim 1- 100$ degenerate pairs in an enlarged parameter
space, it would clearly be preferred over other generic
points.  It is quite interesting that $\tan \beta$ is difficult to
  measure in general, and it is challenging to find signatures that are
  sensitive to $\tan \beta$ alone. Perhaps when information about the
  superpartner spectrum is known from the LHC, $\tan \beta$ could be
  determined with other information that is very sensitive to $\tan
  \beta$, such as the higgs sector, $g_{\mu} -2$ and $B_S
  \rightarrow \mu^+ \mu^-$ constraints.

We have illustrated our approach to the inverse problem in the context of
low-energy SUSY, but the same ideas can be applied to any theory of
physics beyond the standard model, including theories of extra
dimensions with KK parity and of little Higgs models with T-parity,
which have very similar signatures to SUSY.  Indeed, SUSY is likely a
more challenging example due to its large parameter space. It would
be interesting to study the inverse map in other models and between
other models, study how the model footprints differ in signature
space, and test ways of distinguishing qualitatively different
possibilities for new physics at the LHC.  Furthermore, our entire
discussion has been at the electroweak 
  scale in four dimensions. Once the 4D effective lagrangian is
  determined at the weak scale, we can begin probing the underlying
  higher scale or higher dimensional physics.  

\textbf{Acknowledgments:}  We are deeply  indebted to Michael Busha,
Suvendra Dutta, Gus Evrard, and Matias Zaldarriaga for
valuable computing time and computing expertise at two cluster
machines:  ``Sauron'' at the Harvard Center for Astrophysics and
``Opus'' at the Computing for the Natural Sciences
Group, University of Michigan. 
We thank Kevin Black, Matt Bowen, Michael Douglas, Can Kilic, Fabiola
Gianotti, Tao Han, John Huth, Amit Lath, Rakhi Mahbubani, Shawn McKee, Brent
Nelson, Michael Peskin, Aaron Pierce, Albert de Roeck, Maria
Spiropulu, Scott Thomas, Chris Tully, Devin Walker, and James Wells for interesting
discussions.  We especially thank Matt Strassler and Steve Mrenna for
valuable insights into detector simulations and for helping us
incorporate PGS in this study.  The work of NAH, JKT, and LTW is
supported by the DOE under contract DE-FG02-91ER40654. The work of GLK
is supported in part by the DOE.   LTW and GLK thank the Aspen Center for
Theoretical Physics for hospitality while part of this work was
carried out.  LTW also thanks the support and hospitality of the Michigan Center for
Theoretical Physics during several stages of this work.

\appendix

\section{N-tuples and Demonic Bins}
\label{sec:demonic}

In section \ref{sec:degens} we derived a value for $N$, the number of
experimentally distinguishable outcomes at the LHC, making the
assumption that each signature bin is equally likely to be occupied.
We can improve on this very simple picture.   For instance, it could
be that some fraction of bins are 
``demonic", in the sense that a disproportionate number of balls
land into them. We can model this by saying that a fraction $x$ of
balls land into a fraction $\lambda$ of the bins, while the
remaining fraction $(1 - x)$ of the balls land into the $(1 -
\lambda)$ fraction of bins. The number of expected $p$-tuples of
balls in the same bin is then
\begin{equation}
\label{eq:demonic}
N_p \sim \frac{m^p}{p! N^{p - 1}} \left(\frac{x^p}{\lambda^{p - 1}} +
\frac{(1 - x)^p}{(1 - \lambda)^{p-1}} \right).
\end{equation}
We can fit the observed $N_p$ to determine $N,x,\lambda$.

In our sample of with $m = 43026$, we have 283 doubles, 32 triples, and 3 quadruples, which is consistent with
\be
\label{eq:guesstuples}
N = 3.19 \times 10^7, \quad x = 0.0317, \quad \lambda = 0.000114,
\ee
tellling us that 3.2\% of models map to 0.01\% of signature space.
However, with such low statistics it is difficult to say whether this
picture is correct, because we are fitting three numbers with three
parameters.  To really check the picture, we can artificially increase
the statistics by going to a larger signature cut to define degenerate
models.   If we increase the number of doubles by a factor of 20, this
gives us a sizable number of $p$-tuples, and we can now fit the $N_p$
curves as a function of $m$.  Because we have 43,026 models to work
with, we can average over many different subsets containing $m$ models
to get a smooth curve for $N_p$ as a function of $m$. 

\begin{figure}
\begin{center}
\includegraphics[scale=0.9]{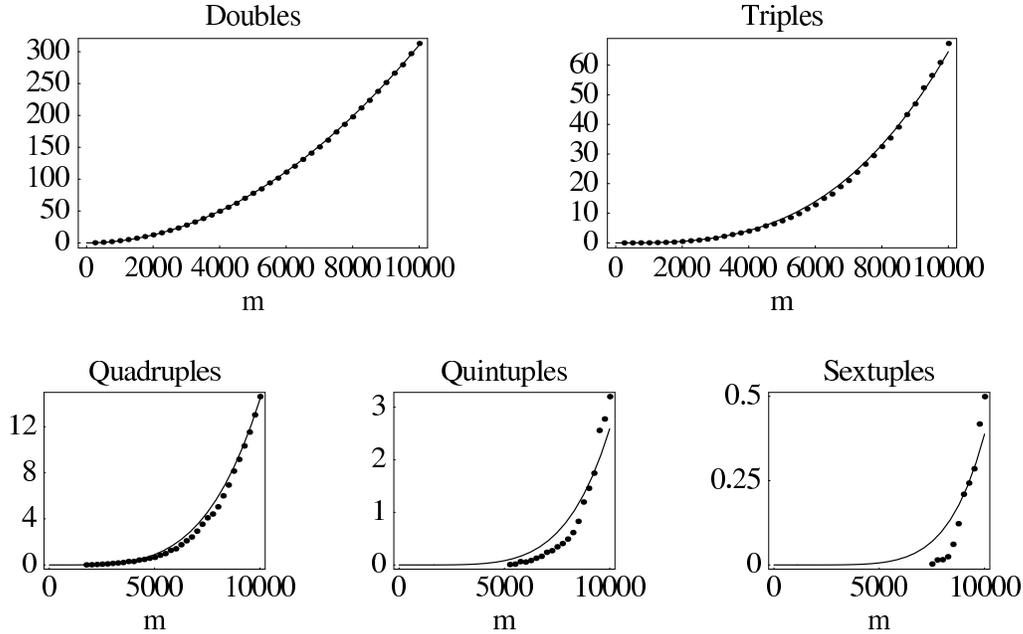}
\end{center}
\caption{The fits of $N_p$ to equation (\ref{eq:demonic}).  The $m$
  axis indicates the number of models simulated, and in order to make
  smooth curves, 1000 random choices for the $m$ models were averaged
  for each data point.   The number of $p$-tuples is consistent with
  the picture that 4.7\% of models map to 0.11\% of signature space,
  indicating that while 5\% of models are much more likely to have
  degeneracies than generic models, the phenomena of degeneracies is
  not completely dominated by small regions in parameter space.} 
\label{fig:tuplefits}
\end{figure}

The curves shown in figure \ref{fig:tuplefits} are for
\be
\label{eq:fittuples}
N = 4.65 \times 10^5, \quad x = 0.0473, \quad \lambda = 0.00113,
\ee
which tells us that 4.7\% of the models map to 0.11\% of the signature
space.  Because we have artificially increased the number of doubles
by a factor of 20, the real value of $N$ should be 20 times bigger
than the value in equation (\ref{eq:fittuples}), giving us 
\be
N = 9.3 \times 10^6,
\ee
which is roughly compatible with equation (\ref{eq:guesstuples}).

\section{Complete List of LHC Observables}
\label{sec:observelist}

In this section, we summarize the LHC signatures we have included in
our study.  The signatures can be divided into two categories: counting
signatures and kinematical distributions.  Both types of signatures depend crucially on the cuts and triggers we impose to isolate a signal sample.

\subsection{Cuts and Triggers}

We identify our signal sample through a set of event filtering triggers and selection cuts.   To account for the effect of multiple interactions and initial
state radiation (which yield additional soft objects in events), we only keep objects in the event record if 
\begin{equation}
\mbox{photon, leptons: } P_T > 20, \quad  |\eta| <2, \qquad
\mbox{jets: } P_T> 50, \quad |\eta| < 3. 
\end{equation}
We then select events with two or more jets subject to the criteria that
\begin{equation}
\not{\!\!E}_T > 150 \GeV, \qquad H_T > \left\{\begin{array}{ll} 600 \GeV
&  \mbox{0 or 1 lepton in event} \\ 400 \GeV & \mbox{2 or more leptons
  in event}\end{array} \right., 
\end{equation}
where
\begin{equation}
\qquad H_T =  \; \not{\!\!E}_T + \sum_{\rm all\, jets} P^a_T.
\end{equation}
These cuts were not optimized to maximally enhance signal over
standard model background.  However, the number of standard model
events that pass these cuts are reasonable---a 10 fb$^{-1}$ $t\bar{t}$
sample contains around 76000  events that satisfy the above criteria out of $4.9 \times 10^6$ events
total. 

\subsection{Counting Signatures}

Counting signatures record the number of a specific kind of events
based solely on their object content.  The most inclusive counting signature is the total number of  events that passes the trigger and cuts. All other counting
signatures are represented as ratios to the total number of
events or as ratios to each other. 

The only distinction between jets is whether they are
$b$-tagged.\footnote{Strictly speaking, the $b$-tag algorithm in PGS is a heavy-flavor tag and thus includes some $c$-quark jets.} Therefore, we count jet multiplicity  in
two categories, $b$-jet and non-$b$-jet.   On the other hand, the
leptons carry both charge and flavor information, so we
keep separate counts for each charge and flavor
configuration. Supersymmetric events could have multiple photons
associated with them both from initial and final state radiation, as
well as decay products.  Therefore, a typical classification of a counting signature could be
represented as 
\bea
\label{count_class}
\mbox{jets}(\mbox{$\#$non-$b$-jet, $\# b$-jet})+\mbox{leptons($\#$,
  charge, flavor)} + \# \gamma + \not{\!\!E}_T.
\eea
Of course, we could form more inclusive observables by combining
several categories together. For example, we could have an inclusive
signature for the number of events with 1 lepton (including all
possible charges and flavors) and 2 non-$b$-jets. 

Now, we list the counting signatures that we have implemented.  Where
appropriate, we use the following shorthand to define ratios. 
\be
\frac{Y}{X} \rightarrow \frac{\mbox{Number of events with X and
    Y}}{\mbox{Number of events with X}} 
\ee

\begin{enumerate}
\item Total number of events. 

The sum of all events that passed the  selection cuts. 

\item Jet and lepton counts.

 Events are categorized
  according to the number of leptonic objects in the event.  All
 counts here are charge and flavor inclusive. For each lepton number
 category, the counts are further separated according to number of
 jets and $b$-jets in the event.   The signatures are:
 \be
 \frac{n_\ell \mbox{ leptons}}{\mbox{total events}}, \qquad  \frac{n_j \mbox{ jets}}{n_\ell \mbox{ leptons}}, \qquad  \frac{n_b \mbox{ $b$-jets}}{n_\ell \mbox{ leptons}},
 \ee
 where
 \be
 n_\ell = 0, 1, 2, 3, 4^+, \qquad n_j = 2,3,4,5,6,7,8^+, \qquad n_b = 0,1,2,3,4,5,6^+.
 \ee
 Note that $n_j \mbox{ jets}$ includes both $b$-jets and non-$b$-jets.

\item  Lepton flavor and charge counts.

For events with leptons, flavor information is used to further divide
signatures into sub-categories. We have included here only the overall
counting of the different flavor/charge categories without regard for
jet multiplicity.  In all cases, the signature used is
\be
\frac{\mbox{flavor category}}{n_\ell \mbox{ leptons}},
\ee
where the flavor categories are:
\begin{itemize}
\item $1\ell$. 6 possibilities $\ell_i^{\pm}$ ($i=e, \mu ,\tau$).
\item $2 \ell$. 21 combinations $\ell_i^{\pm} \ell_j^{\pm}$.
\item $3 \ell$. For events with 2 same flavor opposite charge leptons
  $\ell^+_i \ell^-_i \ell_j^\pm$, we categorize them according to the flavor
  of $\ell_i$ and the flavor/charge of $\ell_j^\pm$.  (18 combinations.)  For trilepton events without same flavor opposite charge pairs, we give the total charge $(1\pm, 3\pm)$ and the dominant flavor ($e, \mu ,\tau$).  (12 combinations.)  When the dominant flavor is ambiguous, we assign the dominant flavor to be $\tau$.
  \item $4^+ \ell$. We do not include flavor/charge information in this
  category.
\end{itemize}
\item Photon counts.
 \be
   \frac{n_\gamma \mbox{ photons}}{\mbox{total events}}, \qquad n_\gamma = 0, 1, 2^+.
  \ee
  
\end{enumerate}

\subsection{Kinematical Distributions}

Generally, separate distributions for certain kinematic variables
can be implemented according to the object content of
the event, as in equation~(\ref{count_class}).  For a lot of inclusive distributions we have used in this
analysis, we use the number of leptons, jets, and $b$-jets to label the distribution. 



The kinematical distributions are divided into two broad categories: $P_T$ (those based on transverse momentum sums)
and $m_{\rm inv}$ (those based on Lorentz-invariant four-vector sums).

\subsubsection{Transverse Momenta:  $P_T$}
In this category, all kinematical observables has the form
\bea
\sum_{\{a \}} P_T^a,
\eea
where the sum is over a set of specific objects of interest.  Distributions are always binned into deciles, and the boundaries of the deciles (excluding the overall lower and upper bounds) are the signatures we store.

The most commonly used observable in this category is effective mass, 
$M_{\rm eff}$ \cite{Baer:1995nq,Baer:1995va,Hinchliffe:1996iu,atlastdr}.
Instead of defining $M_{\rm eff}$ only for the jet 
signatures, we define such an observable for any large class of event
objects, 
\be
M_{\rm {eff}}=\not{\!\!E}_T+\sum_{\rm all \; objects} P_T,
\ee
Notice that instead of summing the $P_T$ of just the jets, we sum the $P_T$
for all objects in the events. We separate $M_{\rm eff}$ into 12 different distributions labeled by
the number of jets and leptons in the event:
\be
n_{j} = 2, 3,4,5^+, \qquad n_\ell = 0, 1,2^+.
\ee
In particular, $M_{\rm eff} (n_j, n_\ell)$ only includes events with exactly $n_j$ jets and $n_\ell$ leptons.

We have also implemented inclusive distributions of jets, leptons, photons, and
$\not{\!\!E}_T$.   
\begin{itemize}
\item $P_T^{nhj}$. Inclusive $P_T$ distribution of the $n$-th hardest
  jet, where $n = 1,2,3,4$. Distributions are categorized according to number of leptons ($n_\ell = 0, 1,2^+$) but include all events with at least $n$ jets.
\item$P_T^{\ell}$ for all events with leptons. Separate distributions
  are used for different lepton flavor, but there is no charge separation, nor separation by the number of jets in the event.
  \item$P_T^{\gamma}$ for all events with photons.
  \item $\not{\!\!E}_T$ for all events that pass our trigger. Separate distributions
  are used for different number of leptons  ($n_\ell = 0, 1,2^+$).
\end{itemize}

\subsubsection{Invariant Mass:  $m_{\rm inv}$}

In this category, all kinematical observables have the form
\bea
\left(\sum_{\{a \}} p_\mu^a \right)^2.
\eea 
Invariant mass distribution typically carry important information
about the masses of the massive particles in the decay chain. They
have been used extensively in the past in studies for measuring soft
parameters \cite{atlastdr, Gjelsten:2004ki,Gjelsten:2005aw,Lester:2005je}. We have implemented in our study a collection of invariant
mass distributions which, if well populated, are sensitive
to all important masses of superpartners.   Unless otherwise indicated, invariant mass distributions are binned into deciles.  
\begin{enumerate}
\item All objects.

Invariant mass for all the objects in the event are
used in there distributions.  Separate distributions are created
according to number of lepton and jets in the event:
\be
n_{j} = 2, 3,4,5^+, \qquad n_\ell = 0, 1,2^+.
\ee

\item Jet objects.

Invariant masses of various combinations of  jets are
used, if such a combination is possible in the event. 
Events are further divided into separate distributions based on the number of leptons in the event $(n_\ell=0,1,2^+)$.  We use the following shorthand notation in events with many jets: $h$ refers to the hardest jet regardless of flavor, $n$ refers to the hardest non-$b$-jet; $b$ refers to the hardest jet with a $b$-tag.   In all but the di-$b$-jet invariant mass distribution, these distributions are inclusive of the total number of jets; for example, the invariant mass distribution for the two hardest jets includes all events with at least two hard jets.

\begin{itemize}
\item $m_{hh}$. Two hardest jets.
\item $m_{hhh}$. Three hardest jets.
\item $m_{hhhh}$. Four hardest jets.
\item $m_{nn}$. Hardest pair of non-$b$-jets.
\item $m_{bn}$. Hardest $b$-jet paired with hardest non-$b$-jet.
\item $m_{bnn}$. Hardest $b$-jet with two hardest  non-$b$-jets.
\item $m_{bb}$. Pairs of $b$-jets in events with exactly 2 $b$-jets, binned in 20-tiles.
\item $m_{bbn}$. Two hardest $b$-jets with hardest non-$b$-jet.
\item $m_{bbnn}$.  Two hardest $b$-jets with 2 hardest non-$b$-jets.
\end{itemize}

\item Events with one lepton.

 For events with a single lepton, we construct invariant mass distributions involving the hardest jets ($h$) and hardest $b$-tagged jets ($b$).  All distributions are separated by lepton flavor but not by charge, and placed into 20-tiles.
 
 \begin{itemize}
 \item $m_{\ell, h^{1,2}}$. Separate distributions for combining
  the single lepton with the first or second hardest jet.
\item $m_{\ell, b^{1,2}}$. Separate distributions for combining a single lepton with the first or second hardest $b$-jet.
\end{itemize}

\item Events with same-flavor opposite-sign dileptons.

For events with dileptons, we have included various invariant
  mass distributions, many of them similar to those studied in
  \cite{Gjelsten:2004ki,Gjelsten:2005aw}. The existence of
  dileptons indicates a long 
  decay chain with at least two jets from earlier stage of the same
  decay chain. Various pairing of the leptons and jets are sensitive
  to a variety of combinations of the masses of the superpartners
  involved---squarks, electroweak-inos, and possibly the sleptons. 

All dilepton invariant mass distributions refer to events
with same-flavor opposite-sign dileptons and no other leptons, with separate distributions
for different dilepton flavors.  All distributions are binned in 20-tiles excepted where noted.
\begin{itemize}
\item $m_{ll}$. Dilepton invariant mass distribution. Dilepton pairs
  whose invariant mass falls in the range $\pm 5$ GeV of $m_Z$ are subtracted and
  histogramed in a separate distribution.  The ratio of the number of events
  in the ``$Z$-window'' to the total number of dileptons is used as an additional signature.
\item $m_{\ell \ell, b}$. Pair of leptons with the hardest $b$-jet in events with at least 1 $b$-jet.
  \item$m_{\ell \ell,hh}$.  Pair of leptons with the two hardest jets.
  \item $m_{\ell \ell, h^{\rm high, low}}$. Combination of
  first or second hardest jet with the pair of leptons. High (low) are
  defined by the combination with the larger (smaller) invariant mass. The
  jet associated with the high (low) combination is labeled as
  high (low) for the distribution below.
\item $m^{{\rm high,low}}(\ell, h^{{\rm high,low}})$. Four
  combinations of combining the high or low jet with each lepton, choosing the lepton that gives
  rise to the larger ($m^{\rm high}$) or smaller ($m^{\rm low}$) invariant mass.
\end{itemize}

The following histogram are for events with exactly 2 $b$-jets and dileptons.   Because these histograms are generically less populated then the ones above, they are binned only in deciles. 

\begin{itemize}
\item $m_{bb}|_{\ell \ell}$  $bb$ invariant mass distribution in
  dilepton events.
\item $m_{bb \ell \ell}$. Pair of leptons with both $b$s.
\item $m_{bb \ell}^{{\rm high, low}}$. $b$-$b$-lepton invariant mass, where high (low) refers to the lepton that gives the largest (smallest) invariant mass.
  \item $m_{\ell \ell, b^{\rm high, low}}$. Combination of
  first or second hardest $b$-jet with the pair of leptons. High (low) are
  defined by the combination with larger (smaller) invariant mass. The
  jet associated with the high (low) combination are labeled as
  high (low) for the distribution below.
\item  $m^{{\rm high,low}}(\ell, b^{{\rm high,low}})$. Four
  combinations of combining the high or low $b$-jet with each lepton, choosing the lepton that gives
  rise to the larger ($m^{\rm high}$) or smaller ($m^{\rm low}$) invariant mass
\end{itemize}

\item {$m_{\gamma \gamma}$}. Di-photon invariant mass distribution, in deciles.
\end{enumerate}

\section{Technique for Assigning Statistical Error Bars}
\label{sec:errors}

To assign error bars to our signatures, we use Gaussian statistics
supplemented with $\mathcal{O}(1)$ scale factors.  We ignore
systematic errors and errors from standard model background
subtraction.  Using a detector simulator, the best way to determine
statistical errors is to run a model many times at fixed integrated
luminosity to figure out the appropriate confidence intervals on a
given signature.  This is very computationally expensive, so we will
use a simple method to mock up reasonable error bars. 

For every signature we first define a raw statistical error bar based on Gaussian statistics.  For an integer-valued counting signature of value $n$,
\begin{equation}
\delta_{\rm raw} n = \sqrt{n+1}.
\end{equation}
For a ratio-valued counting signature of value $n/m$, where $n$ and $m$ are pseudo-independent integer-valued counts,
\begin{equation}
\delta_{\rm raw} \left(\frac{n}{m} \right) = \sqrt{\left(\frac{\sqrt{n+1}}{m+1}  \right)^2 + \left( \frac{n \sqrt{m+1}}{(m+1)^2} \right)^2}.
\end{equation}
For a histogram with $n$ elements and a quantile boundary with value
$q$, we define $q_{\rm low}$ ($q_{\rm high}$) as the new value of the
quantile boundary if $\sqrt{n}$ elements were added to the lowest
(highest) bin of the histogram.  The raw statistical error on $q$ is
defined as 
\begin{equation}
\delta_{\rm raw} q = \sqrt{\frac{( q- q_{\rm low})^2 + (q-q_{\rm high})^2}{2}}.
\end{equation}

While these raw statistical error bars scale like Gaussian statistics,
there is no guarantee that Gaussian statistics is a good description
of the errors from PYTHIA piped through PGS.   Therefore, for the
$i$-th signature, we define the statistical fluctuation as 
\begin{equation}
\delta_{\rm stat} s_i = \alpha_i \delta_{\rm raw} s_i,
\end{equation}
where $\alpha_i$ is an $\mathcal{O}(1)$ phenomenological parameter
that is different for each signature.  To determine the value of
$\alpha$, we ran a subset of our models again with a different random
number seed.  Then for every signature, we calculate 
\begin{equation}
\alpha^{12}_i =  \frac{|s^1_i - s^2_i|}{\sqrt{\left(\delta_{\rm raw} s^1_i\right)^2 + \left(\delta_{\rm raw} s^2_i\right)^2}},
\end{equation}
where $s^1_i$ and $s^2_i$ are the signature values of the same model
with different random number seeds.   The $\alpha_i$ we use in
subsequent calculations is the 95th percentile of $\alpha^{12}_i$
taken over all models or 1, whichever is largest.  The selected
$\alpha_i$ values never exceeded 2.2, telling us that Gaussian
statistics are a good approximation to statistical error bars over a
wide range of models and signatures. 



%
%

%


\small

\begin{thebibliography}{99}

\bibitem{susy}
  S.~Dimopoulos and H.~Georgi,
  Nucl.\ Phys.\ B {\bf 193}, 150 (1981).

\bibitem{technicolor}
  S.~Weinberg,
  Phys.\ Rev.\ D {\bf 19}, 1277 (1979).
  
  L.~Susskind,
  Phys.\ Rev.\ D {\bf 20}, 2619 (1979).


\bibitem{ADD}
  N.~Arkani-Hamed, S.~Dimopoulos and G.~R.~Dvali,
  Phys.\ Lett.\ B {\bf 429}, 263 (1998)
  [arXiv:hep-ph/9803315].
  
N.~Arkani-Hamed, S.~Dimopoulos and G.~R.~Dvali,
  Phys.\ Rev.\ D {\bf 59}, 086004 (1999)
  [arXiv:hep-ph/9807344].
  
  I.~Antoniadis, N.~Arkani-Hamed, S.~Dimopoulos and G.~R.~Dvali,
  Phys.\ Lett.\ B {\bf 436}, 257 (1998)
  [arXiv:hep-ph/9804398].

\bibitem{RS}
  L.~Randall and R.~Sundrum,
  Phys.\ Rev.\ Lett.\  {\bf 83}, 3370 (1999)
  [arXiv:hep-ph/9905221].
  
L.~Randall and R.~Sundrum,
  Phys.\ Rev.\ Lett.\  {\bf 83}, 4690 (1999)
  [arXiv:hep-th/9906064].

\bibitem{Arkani-Hamed:2001nc}
  N.~Arkani-Hamed, A.~G.~Cohen and H.~Georgi,
  Phys.\ Lett.\ B {\bf 513}, 232 (2001)
  [arXiv:hep-ph/0105239].

\bibitem{higgsless}
  C.~Csaki, C.~Grojean, L.~Pilo and J.~Terning,
  Phys.\ Rev.\ Lett.\  {\bf 92}, 101802 (2004)
  [arXiv:hep-ph/0308038].

\bibitem{comphiggs}
  K.~Agashe, R.~Contino and A.~Pomarol,
  Nucl.\ Phys.\ B {\bf 719}, 165 (2005)
  [arXiv:hep-ph/0412089].
\bibitem{warpedgut}
  Z.~Chacko, Y.~Nomura and D.~Tucker-Smith,
  Nucl.\ Phys.\ B {\bf 725}, 207 (2005)
  [arXiv:hep-ph/0504095].

\bibitem{warpedgood}
  K.~Agashe, A.~Delgado, M.~J.~May and R.~Sundrum,
  JHEP {\bf 0308}, 050 (2003)
  [arXiv:hep-ph/0308036].

\bibitem{twinhiggs}
  Z.~Chacko, H.~S.~Goh and R.~Harnik,
  arXiv:hep-ph/0506256.

\bibitem{susylittlehiggs}
  T.~Roy and M.~Schmaltz,
  arXiv:hep-ph/0509357.
  
  C.~Csaki, G.~Marandella, Y.~Shirman and A.~Strumia,
  arXiv:hep-ph/0510294.


\bibitem{split}
  N.~Arkani-Hamed, S.~Dimopoulos, G.~F.~Giudice and A.~Romanino,
  Nucl.\ Phys.\ B {\bf 709}, 3 (2005)
  [arXiv:hep-ph/0409232].
  
  N.~Arkani-Hamed and S.~Dimopoulos,
  JHEP {\bf 0506}, 073 (2005)
  [arXiv:hep-th/0405159].
  
  G.~F.~Giudice and A.~Romanino,
  Nucl.\ Phys.\ B {\bf 699}, 65 (2004)
  [Erratum-ibid.\ B {\bf 706}, 65 (2005)]
  [arXiv:hep-ph/0406088].

\bibitem{finetune}

  N.~Arkani-Hamed, S.~Dimopoulos and S.~Kachru,
  arXiv:hep-th/0501082.

\bibitem{finetune-others}
  P.~C.~Schuster and N.~Toro,
  Phys.\ Rev.\ D {\bf 72}, 093005 (2005)
  [arXiv:hep-ph/0506079].
  L.~Senatore,
  arXiv:hep-ph/0507257.
  R.~Mahbubani and L.~Senatore,
  arXiv:hep-ph/0510064.


\bibitem{Appelquist:2000nn}
  T.~Appelquist, H.~C.~Cheng and B.~A.~Dobrescu,
  Phys.\ Rev.\ D {\bf 64}, 035002 (2001)
  [arXiv:hep-ph/0012100].

\bibitem{Cheng:2003ju}
  H.~C.~Cheng and I.~Low,
  JHEP {\bf 0309}, 051 (2003)
  [arXiv:hep-ph/0308199].



\bibitem{Cheng:2002ab}
  H.~C.~Cheng, K.~T.~Matchev and M.~Schmaltz,
  Phys.\ Rev.\ D {\bf 66}, 056006 (2002)
  [arXiv:hep-ph/0205314].

\bibitem{Barr:2004ze}
  A.~J.~Barr,
  Phys.\ Lett.\ B {\bf 596}, 205 (2004)
  [arXiv:hep-ph/0405052].

\bibitem{Goto:2004cp}
  T.~Goto,
  arXiv:hep-ph/0411360.



\bibitem{Smillie:2005ar}
  J.~M.~Smillie and B.~R.~Webber,
  JHEP {\bf 0510}, 069 (2005)
  [arXiv:hep-ph/0507170].

\bibitem{Cheng:2005as}
  H.~C.~Cheng, I.~Low and L.~T.~Wang,
  arXiv:hep-ph/0510225.

\bibitem{Datta:2005vx}
  A.~Datta, G.~L.~Kane and M.~Toharia,
  arXiv:hep-ph/0510204.

\bibitem{Datta:2005zs}
  A.~Datta, K.~Kong and K.~T.~Matchev,
  arXiv:hep-ph/0509246.

\bibitem{Barr:2005dz}
  A.~J.~Barr,
  arXiv:hep-ph/0511115.


\bibitem{Binetruy:2003cy}
  P.~Binetruy, G.~L.~Kane, B.~D.~Nelson, L.~T.~Wang and T.~T.~Wang,
  Phys.\ Rev.\ D {\bf 70}, 095006 (2004)
  [arXiv:hep-ph/0312248].


\bibitem{Kane:2005az}
  G.~Kane,
  arXiv:hep-ph/0504257.


\bibitem{Baer:1995nq}
  H.~Baer, C.~h.~Chen, F.~Paige and X.~Tata,
  Phys.\ Rev.\ D {\bf 52}, 2746 (1995)
  [arXiv:hep-ph/9503271].

\bibitem{Baer:1995va}
  H.~Baer, C.~h.~Chen, F.~Paige and X.~Tata,
  Phys.\ Rev.\ D {\bf 53}, 6241 (1996)
  [arXiv:hep-ph/9512383].

\bibitem{Mrenna:1995ax}
  S.~Mrenna, G.~L.~Kane, G.~D.~Kribs and J.~D.~Wells,
  Phys.\ Rev.\ D {\bf 53}, 1168 (1996)
  [arXiv:hep-ph/9505245].

\bibitem{gmsb}
  S.~Dimopoulos, S.~D.~Thomas and J.~D.~Wells,
  Nucl.\ Phys.\ B {\bf 488}, 39 (1997)
  [arXiv:hep-ph/9609434].

\bibitem{Hinchliffe:1996iu}
  I.~Hinchliffe, F.~E.~Paige, M.~D.~Shapiro, J.~Soderqvist and W.~Yao,
  Phys.\ Rev.\ D {\bf 55}, 5520 (1997)
  [arXiv:hep-ph/9610544].

\bibitem{amsb}
  T.~Gherghetta, G.~F.~Giudice and J.~D.~Wells,
  Nucl.\ Phys.\ B {\bf 559}, 27 (1999)
  [arXiv:hep-ph/9904378].

\bibitem{atlastdr}
For a summary of earlier studies, ATLAS Detector and Physics
Performance Technical Design Report, Chapter 20 \\
http://atlas.web.cern.ch/Atlas/GROUPS/PHYSICS/TDR/access.html


\bibitem{Gjelsten:2004ki}
  B.~K.~Gjelsten, D.~J.~Miller and P.~Osland,
  JHEP {\bf 0412}, 003 (2004)
  [arXiv:hep-ph/0410303].

\bibitem{Gjelsten:2005aw}
  B.~K.~Gjelsten, D.~J.~Miller and P.~Osland,
  JHEP {\bf 0506}, 015 (2005)
  [arXiv:hep-ph/0501033].

\bibitem{Gjelsten:2005vv}
  B.~K.~Gjelsten, D.~J.~Miller and P.~Osland,
  arXiv:hep-ph/0511008.

\bibitem{Lester:2005je}
  C.~G.~Lester, M.~A.~Parker and M.~J.~White,
  arXiv:hep-ph/0508143.

\bibitem{Lafaye:2004cn}
  R.~Lafaye, T.~Plehn and D.~Zerwas,
  arXiv:hep-ph/0404282.
\bibitem{Bechtle:2005qj}
  P.~Bechtle, K.~Desch and P.~Wienemann,
  arXiv:hep-ph/0511137.
\bibitem{Weiglein:2004hn}
  G.~Weiglein {\it et al.}  [LHC/LC Study Group],
  arXiv:hep-ph/0410364.


\bibitem{pythia}
  T.~Sjostrand, P.~Eden, C.~Friberg, L.~Lonnblad, G.~Miu, S.~Mrenna
  and E.~Norrbin,
  Comput.\ Phys.\ Commun.\  {\bf 135}, 238 (2001)
  [arXiv:hep-ph/0010017].
http://www.thep.lu.se/$\sim$torbjorn/Pythia.html

\bibitem{pgs}
PGS, Simple simulation package for generic collider detectors \\
http://www.physics.ucdavis.edu/$\sim$conway/research/software/pgs/pgs.html

\bibitem{Hinchliffe:1998ys}
  I.~Hinchliffe and F.~E.~Paige,
  Phys.\ Rev.\ D {\bf 60}, 095002 (1999)
  [arXiv:hep-ph/9812233].

\bibitem{Paige:1999ui}
  F.~E.~Paige and J.~D.~Wells,
  arXiv:hep-ph/0001249.


\bibitem{jetcharge}
  K.~Abe {\it et al.}  [SLD Collaboration],
  Phys.\ Rev.\ Lett.\  {\bf 74}, 2890 (1995). 
  
  D.~Buskulic {\it et al.}  [ALEPH Collaboration],
  Z.\ Phys.\ C {\bf 71}, 357 (1996). 
  
  P.~Abreu {\it et al.}  [DELPHI Collaboration], 
  Z.\ Phys.\ C {\bf 72}, 17 (1996). 
  
  R.~Akers {\it et al.}  [OPAL Collaboration],
  Phys.\ Lett.\ B {\bf 327}, 411 (1994). 
  
  F.~Abe {\it et al.}  [CDF Collaboration],
  Phys.\ Rev.\ D {\bf 60}, 072003 (1999)
  [arXiv:hep-ex/9903011]. 
  
  X.~Zhang,
UMI-31-34389. 

  P.~Eerola  [ATLAS Collaboration],
  Nucl.\ Instrum.\ Meth.\ A {\bf 384}, 93 (1996)
  [arXiv:hep-ex/9610002].


\bibitem{sps}
  B.~C.~Allanach {\it et al.},
in {\it Proc. of the APS/DPF/DPB Summer Study on the Future of
  Particle Physics (Snowmass 2001) } ed. N.~Graf, 
  Eur.\ Phys.\ J.\ C {\bf 25}, 113 (2002)
  [eConf {\bf C010630}, P125 (2001)]
  [arXiv:hep-ph/0202233].

\bibitem{Allanach:2004my}
  B.~C.~Allanach, D.~Grellscheid and F.~Quevedo,
  JHEP {\bf 0407}, 069 (2004)
  [arXiv:hep-ph/0406277].



\end{thebibliography}
\end{document}